\documentclass[%
 reprint,
 superscriptaddress,
 floatfix,
 amsmath, amssymb,
 aps,
 pra,
longbibliography]{revtex4-2}

\usepackage[utf8]{inputenc}
\usepackage{enumerate}
\usepackage{graphicx}
\usepackage{caption}
\usepackage{subcaption}

\usepackage{dcolumn}
\usepackage{bm}
\usepackage[unicode, colorlinks=true,linkcolor=blue,citecolor=blue,urlcolor=blue]{hyperref}
\usepackage{amsthm}

\usepackage{changes}

\usepackage[justification=justified,
   format=plain]{caption} 

\usepackage{xcolor}
\usepackage{braket}
\usepackage{float}
 \usepackage{algorithm}
\usepackage[noend]{algpseudocode}
\usepackage{tikz}
\usetikzlibrary{positioning, calc}
\usepackage{ragged2e}

\usepackage{bbold}

\usepackage{calc}
\usepackage{accents}

\newtheorem{theorem}{Theorem}[section]
\renewcommand{\arraystretch}{1.6}
\usepackage{booktabs}
\usepackage{soul}

\begin{document}

\preprint{APS/123-QED}

\title{Boosting projective methods for quantum process and detector tomography}

\author{Júlia Barberà-Rodríguez}
\email{julia.barbera@icfo.eu}
\affiliation{ICFO - Institut de Ciencies Fotoniques, The Barcelona Institute of Science and Technology, Castelldefels (Barcelona) 08860, Spain}
\author{Leonardo Zambrano}
\affiliation{ICFO - Institut de Ciencies Fotoniques, The Barcelona Institute of Science and Technology, Castelldefels (Barcelona) 08860, Spain}
\author{Antonio Acín}
\affiliation{ICFO - Institut de Ciencies Fotoniques, The Barcelona Institute of Science and Technology, Castelldefels (Barcelona) 08860, Spain}
\affiliation{ICREA-Institucio Catalana de Recerca i Estudis Avançats, Lluis Companys 23, 08010 Barcelona, Spain}

\author{Donato Farina}
\affiliation{Physics Department E. Pancini - Università degli Studi di Napoli Federico II, Complesso Universitario Monte S. Angelo - Via Cintia - I-80126 Napoli, Italy}

\date{\today}

\begin{abstract}

We introduce two methods for quantum process and detector tomography. In the quantum process tomography method, we develop an analytical procedure for projecting the linear inversion estimation of a quantum channel onto the set of completely positive trace-preserving matrices. By integrating this method with alternate projection techniques, we achieve a three-order-of-magnitude improvement in approximating the closest quantum channel to an arbitrary Hermitian matrix compared to existing methods without compromising computational efficiency.
Our second method extends this approach to quantum detector tomography, demonstrating superior efficiency compared to current techniques. Through numerical simulations, we evaluate our protocols across channels of up to four qubits in quantum process tomography and systems of up to six qubits in quantum detector tomography, showcasing superior precision and efficiency.


\end{abstract}

\maketitle

\section{Introduction}
%
%
%
The advancement of quantum technologies passes through the development of methods that are capable of precisely and efficiently reconstructing quantum states and quantum channels from quantum measurements.
Achieving this demands a precise understanding of the measurements in use, necessitating a thorough characterization of detectors. Consequently, disciplines such as quantum state tomography (QST)~\cite{james2001measurement, thew2002qudit}, quantum process tomography (QPT)~\cite{chuang1997prescription, poyatos1997complete, mohseni2008quantum} and detector tomography (QDT)~\cite{luis1999complete, fiuravsek2001maximum, d2004quantum, lundeen2009tomography} stand as interconnected and indispensable tasks within the field of quantum information science.

The simplest tomographic methods deliver an estimator for a state, process or detector as the solution of a system of linear equations.  A central challenge in this context is that the mathematical object reconstructed via linear inversion often does not correspond to a physical object. This issue arises primarily due to the finite nature of the measured data. Consequently, when confronted with a nonphysical reconstruction as input, it becomes crucial to determine a proper way of associating a physical object to it.

This task is commonly addressed by minimizing suitable cost functions.
A common choice is to determine a physical matrix that  minimizes the Frobenius distance from the input matrix. 
Geometrically, this procedure identifies the (unique) Euclidean projection of the input matrix onto the convex set of physical matrices, the specifics of which depend on the problem at hand.
Interestingly,
for QST (and ancilla-assisted QPT) under some hypotheses ~\cite{smolin2012efficient}, this is equivalent to finding a maximum likelihood matrix that is consistent with the physical requirements. 
%
A major issue is that, when approached with brute-force convex optimization methods, the complexity of the postprocessing step involving the projection is subject to exponential scaling in the system size, typical in quantum mechanical problems. 
This underscores the importance of seeking efficient methods for this task in the context of quantum tomography of states, channels, and detectors.
Interestingly, an analytic solution for the projection exists in the case of quantum states~\cite{verstraete2002geometry, smolin2012efficient}. While it does not resolve the exponential scaling issue, it offers a significant advantage over brute-force convex optimization approaches~\cite{smolin2012efficient}. On the contrary, the fields of quantum process tomography and detector tomography still lack of analytic projection methods and recently are attracting increasing interest~\cite{knee2018quantum,Surawy_Stepney_2022,quiroga2023using,di2024fourier,bao2023ancilla,jaouni2024quantum, xiao2024two, wang2021two,pokharel2024scalable}.
Additionally, it is worth mentioning that projective approaches are also useful in other contexts, for instance to restore formal properties of relevant master equations 
while retaining non-Markovian effects~\cite{PhysRevX.14.031010}.

Our work focuses on  developing precise and efficient projection techniques for direct and ancilla-assisted process tomography and positive operator-valued measure (POVM) tomography. Given a non-physical estimate, our approach focuses on the algorithmic projection of the object onto the valid set of quantum channels or POVMs, ensuring it satisfies the necessary physical constraints. For QPT, we present an analytical approximate projection scheme that, given an arbitrary, non-physical estimation of the quantum process as input, finds a completely positive (CP) and trace-preserving (TP) matrix that is close to the exact projection onto the $CPTP$ set, that is, the set of CPTP linear maps. This method, when combined with alternate projection techniques, enables us to achieve more precise solutions compared to state-of-the-art methods. The method we propose is similar in spirit to the iterative projection in Ref.~\cite{knee2018quantum}, but differs by projecting onto the set of density matrices instead of the positive set. We enhance the algorithm with a final correction step, based on the Cholesky decomposition, referred to as the CBA correction, providing more precise solutions. Furthermore, we adapt the calculations to tackle the problem of QDT, developing an analytical approximate projection method that surpasses recent methods in precision.

The article is organized as follows.
In Sec.~\ref{sec:prel} we introduce the formalism and the notation for QPT and QDT.
Our original techniques are described in
Sec.~\ref{sec:original}, together with numerical tests, to finally conclude in Sec.~\ref{sec:conc}.

\section{Preliminaries}
\label{sec:prel}
We start by reviewing the conditions that quantum channels and POVMs must satisfy, as well as the projection of arbitrary matrices onto the sets of quantum channels and measurements.

\subsection{Quantum channels}
%
%
Let $\mathfrak{L}(\mathbb{C}^d)$ denote the space of linear operators associated with a finite-dimensional quantum system $S$ of dimension $d$. The physical evolution of quantum states within $S$ is described by a quantum channel $\Phi: \mathfrak{L}(\mathbb{C}^d) \rightarrow \mathfrak{L}(\mathbb{C}^d)$, a linear super-operator mapping density matrices to density matrices~\cite{wilde2013quantum, watrous2018theory}. To ensure the validity of this description, $\Phi$ must satisfy the CPTP conditions.

A convenient representation of a quantum channel $\Phi$ is provided by its Choi matrix. The Choi matrix $J(\Phi)$ of $\Phi$ is constructed by considering a maximally entangled state between the system $S$ and a duplicate system $A$, upon which $\Phi$ acts locally on $S$:
\begin{align}
J_{}=({\Phi} \otimes \mathbb{1}_d) \,\left(\frac{1}{d} \sum_{i,j=1}^d \ket{i}\bra{j}_S\otimes \ket{i}\bra{j}_A \right),
    \label{choi-state}
\end{align}
where ${\ket{i}}_{i=1}^d$ represents an orthonormal basis of $\mathbb{C}^d$, and $\mathbb{1}_d$ denotes the identity matrix on $\mathbb{C}^d$.

A Choi matrix $J_{}$ is inherently positive-semidefinite (PSD) with unit trace, thereby representing a quantum state associated with a bipartite system $SA$ of dimension $d^2$. 
In this representation, the TP condition for the channel $\Phi$ is guaranteed by the property that the reduced density matrix of $J$ for subsystem $A$ is the maximally mixed state,
\begin{align}
{\rm tr}_S\, \left(  J_{} \right) = \frac{\mathbb{1}_d}{d}.
    \label{trace preserving}
\end{align}
Conversely, any PSD matrix acting on the bipartite system $SA$ that satisfies Eq.~\eqref{trace preserving} is a valid Choi matrix, in the sense that it corresponds to an associated channel acting on $S$. In fact, from the Choi matrix, one can compute the action of $\Phi$ on any quantum state within $S$, formally 
\begin{equation}
    {\Phi} [\rho_S]=d\,  {\rm tr}_A \left( \left(\mathbb{1}_d \otimes \rho_A^T \right)J_{} \right).
    \label{choi-rhoS-t}
\end{equation}
Here, $\rho_A^T$ represents the transposed state of $\rho_S$, which is initially considered for system $S$ but is then reassigned to system $A$.
\subsection{Quantum process tomography}
Following Ref.~\cite{Nielsen2021gatesettomography}, we provide a brief introduction to quantum process tomography based on linear inversion.

Quantum process tomography aims to characterize an unknown quantum process $\Phi$ by measuring an ensemble of independent copies of quantum states.
In principle, by preparing a maximally entangled state and applying the channel locally on one part, one can obtain the Choi matrix, Eq.~\eqref{choi-state}. Performing a QST of this state is a first way to obtain a representation of the channel, which serves as input for our routines.
This approach is referred as ancilla-assisted process tomography. However, there exist other approaches not requiring the preparation of the aforementioned maximally entangled state which is often complicated in practice.

 To achieve this, we begin with an informationally complete set of input states $\{\rho_i\}$, which serves as a fixed reference basis. We prepare multiple copies of each input state and then evolve them through the process we intend to reconstruct, $\Phi$.
After this evolution, the states are measured using a known informationally complete POVM $\{F_n\}$. Consequently, when measuring a state ${\Phi} [\rho_i]$ using $\{F_n\}$, the probability of observing the outcome $n$ is given by  

\begin{equation}
    p_{n,i} = {\rm tr}(F_n {\Phi} [\rho_i]) = \langle\langle F_n| X|\rho_i\rangle\rangle,
\end{equation}
where $X$ is the natural representation of the channel and $|O\rangle\rangle$ is the vectorized representation of an operator $O$~\cite{watrous2018theory}. We define the matrix $M$ to represent the known measurement effects $M = (\langle\langle F_1|,...,\langle\langle F_{d^2}|)^T$ in a vectorized form, and the input states used in the experiment by the matrix $N = (|\rho_1\rangle\rangle,...,|\rho_{d^2}\rangle\rangle)$. Then, the probability matrix $P$ with entries, $p_{n,i}$, can be written as
\begin{equation}
    P = MX N.
\end{equation}
This allows to reconstruct the process via linear inversion
\begin{equation}
    X = M^{-1}PN^{-1}.
\end{equation}
The matrix $X$ is not exactly the Choi matrix of the channel, but its coefficients can be reordered and normalized to obtain the latter~\cite{watrous2018theory}. However, linear inversion often results in estimates that do not satisfy physical constraints. These violations can be corrected by projecting the non-physical estimate onto the set of valid quantum channels. In this work, we focus on developing this projection algorithm to refine such estimate. Although other techniques~\cite{fiuravsek2001maximum} can produce valid quantum channels, they become inefficient for large systems.

\subsection{Projection onto quantum channels}

\begin{figure}[t]
    \includegraphics[width=1.0\columnwidth]{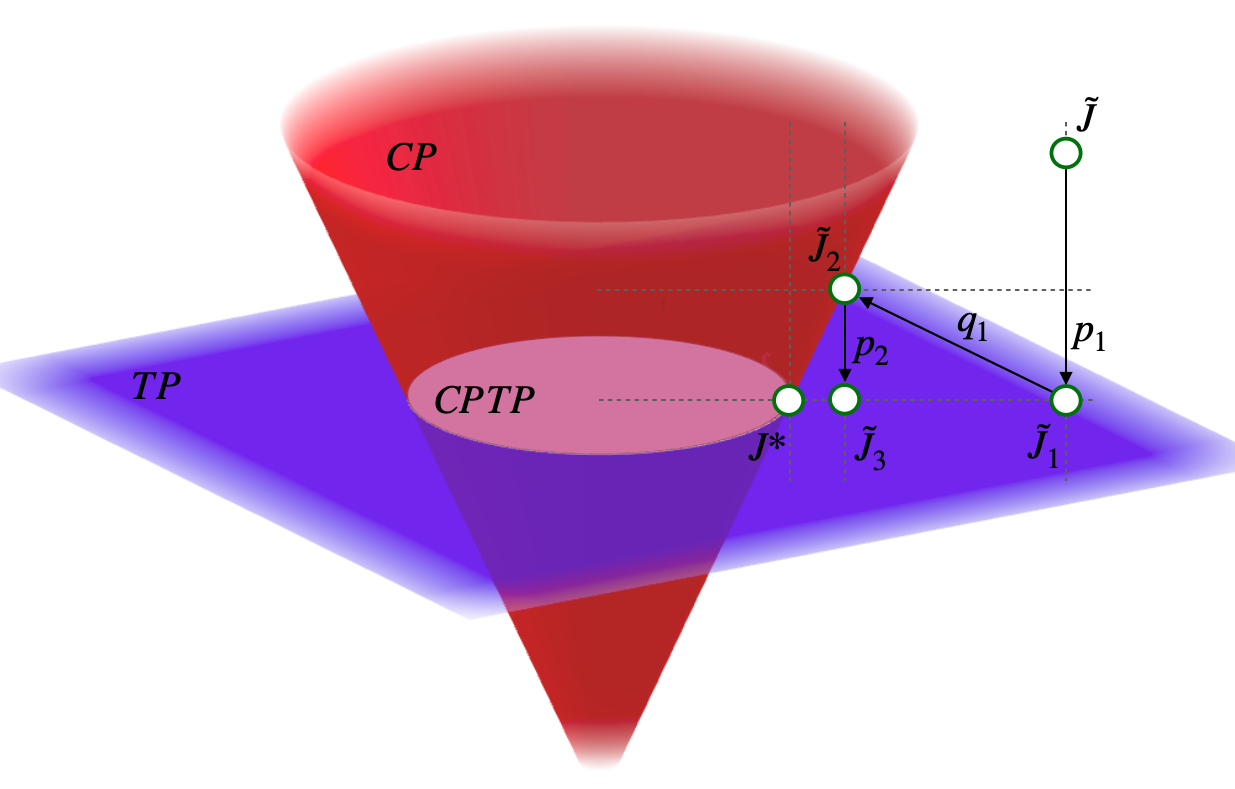}
    \caption{\justifying The red cone denoted as $CP$ contains the set of density matrices $CP1$ in $\mathfrak{L}(\mathbb{C}^{d^2})$ that represent completely positive maps. The set of trace-preserving Hermitian matrices forms the blue hyperplane, whose elements are described by $d^4 -d^2$ independent real parameters. Their intersection gives rise to the set of Choi matrices. In the figure, we also illustrate how to address the QPT problem using Dykstra's projection algorithm when dealing with a noisy Choi state $\tilde{J}$ as input.}
    \label{fig:cptp-projection}
\end{figure}

The task of QPT is to estimate an unknown quantum channel acting on a quantum system based on measurement data obtained from the system. The simplest method for reconstruction involves solving a system of linear equations to obtain an estimate $\tilde{J}$ of the original Choi matrix $J$ representing the process~\cite{chuang1997prescription, poyatos1997complete}. However, due to the inherent limitations of the reconstruction process and finite statistics effects, this estimation often fails to maintain the CPTP properties. To ensure the physical validity of the solution, it becomes necessary to project $\tilde{J}$ onto the set of quantum channels (depicted in Fig.\,\ref{fig:cptp-projection}). This projection can be formulated as the following minimization problem over the Frobenius norm, defined as $||A - B||_F^2 = \text{Tr}[(A - B)^\dagger(A - B)]$,
\begin{align}\label{eq:proj_cptp}
    J^* =  \mathrm{arg} \min_{Z} & \; \lVert \tilde{J} - Z  \rVert_F^2  \nonumber \\
\textrm{s.t.}  &  \quad Z \geq 0  \\
& \quad {\rm tr}_S\,  Z = \frac{\mathbb{1}_d}{d}. \nonumber
\end{align}
These constraints ensure that the optimization is performed over all $d^2 \times d^2$ Hermitian matrices $Z$ that are PSD and TP.
Geometrically, $\lVert \tilde{J} - J^*  \rVert_F $ is the distance of $\tilde J$ from the set of Choi states.

A possible way of solving problem~\eqref{eq:proj_cptp} and finding the projection is to use semidefinite programming (SDP) routines. For multipartite systems, $d$ scales exponentially with the number of qubits in the system, and since one has to optimize over $d^4$ variables, the optimization problem becomes impossible to solve in reasonable time scales.

\subsection{POVMs}
%
In quantum mechanics, measurements are often described using the POVM formalism. For a quantum system of dimension $d$, an $N$-outcome POVM is defined as a set $\{ F_n \}_{n=1}^N$ of operators in $\mathfrak{L}(\mathbb{C}^d)$ that satisfy the conditions of summation to identity and positivity, that is,
\begin{align}\label{eq:conditions_povm}
    \sum_{n=1}^N  F_n = \mathbb{1} \qquad \textrm{and} \qquad F_n \geq 0\,,
\end{align}
for all $n=1, 2, ..., N$. When a quantum state $\rho$ is measured using the POVM $\{F_n \}_{n=1}^N$, the probability of obtaining outcome $n$ is given by $p_n = \text{tr}(\rho F_n)$. The conditions in Eq.~\eqref{eq:conditions_povm} ensure that the probabilities are non-negative and normalized.
\subsection{Quantum detector tomography}
As for QPT, we follow Ref.~\cite{Nielsen2021gatesettomography}, to provide a brief overview of quantum detector tomography.
\\
While quantum process tomography focuses on characterizing an unknown quantum process by measuring how different input states evolve, quantum detector tomography aims to reconstruct the POVMs that define the measurement outcomes.

In QDT, we similarly prepare a known informationally complete set of input states $\{\rho_i\}$, but instead of evolving them through a process, these states are directly measured by the POVMs $\{F_n\}$ we aim to characterize. The probability of obtaining outcome $F_n$ when measuring $\rho_i$ is given by $\boldsymbol{p}_{n,i} = {\rm tr}(\rho_i F_n)$.
Given this, and using the same matrix $N$ from QPT that encodes the input states, along with the probability matrix $P$, we can reconstruct the desired POVM elements encoded in $M$ via linear inversion as follows

\begin{equation}
    M = P N^{-1}.
\end{equation}

This method allows for efficient reconstruction of the POVM, though, as with QPT, the resulting operators may require projection onto the physical set of valid POVMs to ensure positivity and completeness.

\subsection{Projection onto the POVM set}
%

Similar to quantum process tomography, we can also estimate the POVM associated with a measurement process using experimental data from the system. Linear inversion tomography of a POVM $\{F_n\}_{n=1}^N$ typically yields a set of matrices $\{\tilde{F}_n\}_{n=1}^N$ that often do not satisfy the POVM conditions. Therefore, we seek to operate on this set to find the closest valid POVM. This leads us to solve the following optimization problem
\begin{align}\label{eq:proj_povm}
    \{Z_{n}^{*}\} = \mathrm{arg}\min_{ \{Z_n\} } & \; \sum_{n=1}^N \lVert \tilde{F}_n - Z_n \rVert_F^2 \nonumber\\
\textrm{s.t.}  & \quad Z_n \geq 0\\
& \quad \sum_{n=1}^N Z_n = \mathbb{1}. \nonumber
\end{align}
Analogously to the projection onto the set of quantum channels, this problem can be tackled using convex optimization routines which become increasingly challenging for many-particle systems.

\section{Methods and results}
\label{sec:original}

Finding exact solutions to problems \eqref{eq:proj_cptp} and \eqref{eq:proj_povm} turns exceptionally demanding for systems with more than a few qubits. Therefore, our objective is to develop algorithms that offer close approximations to these problems, prioritizing computational efficiency while maintaining a high level of accuracy. In this section, we introduce an approach for approximately projecting onto the set of quantum channels, along with an extension of this method for approximating projections onto the set of POVMs.

\subsection{Quantum Process Tomography with Cholesky-based approximation}

 Here, we introduce an algorithm designed to efficiently identify a Choi state that closely approximates the exact solution while maintaining good computational performance. We call this method the Cholesky-based approximation (CBA). In our approach, we aim to establish an upper bound on the exact solution
 $\lVert \tilde{J} - J^*  \rVert_F^2 $ by replacing the original problem~\eqref{eq:proj_cptp} into two related ones that can be solved analytically:
\begin{enumerate}[(i)]
    \item Given an arbitrary non-physical estimation $\tilde{J}$ of a Choi matrix $J$  acting on a bipartite system $SA$, our first step is to identify its nearest density matrix, that is, a projection onto the $CP1$ set. This involves solving the optimization problem
    \begin{align}
    X^{*} = \mathrm{arg} \min_{X} & \; \lVert \tilde{J} - X \rVert_F^2 \nonumber\\
    \textrm{s.t.} & \quad X \geq 0\\
    & \quad {\rm tr}\,  \left(X\right) = 1.\nonumber
    \end{align}
    \item Then, we can express $X^{*}$ in its Cholesky decomposition, $X^{*} = E E^\dagger$, with $E$ a lower triangular complex matrix, and solve the problem 
    \begin{align}
    Y^{*} = \mathrm{arg} \min_{Y} &\; \lVert E - Y \rVert_F^2 \nonumber\\
    \textrm{s.t.} & \quad {\rm tr}_S\,  \left( YY^\dagger \right) = \frac{\mathbb{1}_d}{d}. 
    \end{align}
\end{enumerate}

Thus, an approximate solution for the closest Choi matrix to the matrix $J$ can be obtained as $Y^{*} {Y^{*}}^\dagger$. In step (i) of our procedure, we introduce the Cholesky decomposition into the problem, allowing us to operate within the Cholesky matrix space. Consequently, in step (ii), the positivity constraint becomes unnecessary, as our variable $Y$ is already within this transformed space. Furthermore,
we simplify the problem described in (ii) by replacing the minimization of $\lVert E E^\dagger - Y Y^\dagger \rVert_F^2$ with the minimization of $\lVert E - Y \rVert_F^2$.
The solution $Y^*$ to this problem will return a Choi matrix
$Y^* Y^*{}^\dagger$.
Since the exact projection $J^*$ of \eqref{eq:proj_cptp} is, by definition, the closest Choi state to the input matrix $\tilde J$ (put differently, any other Choi state will be farther or equally distant),
we necessarily have
$\lVert \tilde J - Y^* Y^*{}^\dagger \rVert_F^2 \geq \lVert \tilde J - J^* \rVert_F^2$.
Hence, the routine will produce an upper bound on the exact distance of $\tilde J$ from the Choi-state set.

For a matrix $\tilde{J}$ acting on $SA$, the solution to problem (i) can be determined analytically using the algorithm presented in Ref.~\cite{smolin2012efficient}. To find an analytical solution to step (ii) we employ Lagrange multipliers and express the constrained problem as an unconstrained one using the Lagrangian
\begin{align}
    \mathcal{L} = \lVert E - Y \rVert_F^2 - \mathrm{tr} \left(\left( \mathbb{1}_d \otimes \Lambda \right) \left(YY^\dagger - \frac{\mathbb{1}_d \otimes \mathbb{1}_d}{d^2} \right)\right),
\end{align}
where $\Lambda$ is a $d \times d$ Hermitian matrix that acts on system $A$, and $\mathbb{1}_d$ is a $d \times d$ identity matrix. The solution $Y$ can be computed by solving the following matrix equation:
\begin{align}\label{eq:derivative_lagrangian_choi}
    \frac{\partial \mathcal{L}}{\partial \overline{Y}} = Y - E -  \left( \mathbb{1}_d \otimes \Lambda \right)\, Y = 0,
\end{align}
where we calculate the matrix derivatives using the results of Wirtinger calculus from Ref.~\cite{koor2023short}. From Eq.~\eqref{eq:derivative_lagrangian_choi}, we get
\begin{align}
    Y =  \mathbb{1}_d \otimes(\mathbb{1}_d - \Lambda)^{-1} E. 
\end{align}
Since the $TP$ constraint ${\rm tr}_S\,  \left( YY^\dagger \right) = \mathbb{1}_d/{d}$ must be satisfied, the value for the Lagrange multipliers $\Lambda$ can be found by writing
\begin{align}
    \frac{\mathbb{1}_d}{d} &= \mathrm{tr}_S \left(Y Y^\dagger \right)  \nonumber \\ 
        &= \mathrm{tr}_S\left(\left( \mathbb{1}_d \otimes(\mathbb{1}_d -  \Lambda)^{-1} \right) EE^\dagger \left(  \mathbb{1}_d \otimes(\mathbb{1}_d -  \Lambda)^{-1}\right) \right) \nonumber \\ 
    &=(\mathbb{1}_d - \Lambda)^{-1} \mathrm{tr}_S\left( EE^\dagger \right) (\mathbb{1}_d - \Lambda)^{-1},
\end{align}
which results in
\begin{align}
    (\mathbb{1}_d - \Lambda)^{2} = d \,\mathrm{tr}_S\left(  E E^\dagger \right).
\end{align}
This implies that the solution to the closest Choi matrix problem using the CBA method is given by 
\begin{align}\label{eq:closest_choi_cholesky}
{Y^*}{Y^*}^\dagger = \left( \mathbb{1}_d 
 \otimes \mathrm{tr}_S(EE^\dagger)^{-\frac{1}{2}} \right) \frac{EE^\dagger}{d} \left(\mathbb{1}_d 
 \otimes \mathrm{tr}_S(EE^\dagger)^{-\frac{1}{2}}\right),
\end{align}
with $E E^\dagger = X^{*}$. 

The solution of step (ii), given by Eq.~\eqref{eq:closest_choi_cholesky}, is equivalent to the one found in Ref.~\cite{xiao2022two}, although a different analytical approach and motivation are employed in the calculation of the projection.

The approximate projection procedure above can be enhanced by combining it with an established algorithm designed to find the closest Choi matrix. This approach involves Dykstra's projection algorithm (refer to App.~\ref{appendix:dykstra} for details). Ideally, the application of Dykstra's 
projection algorithm (depicted in Fig.\,\ref{fig:cptp-projection}) alone should solve problem \eqref{eq:proj_cptp}~\cite{knee2018quantum}. 
However, the convergence of this method is only asymptotic. For a small number of iterations, Dykstra's algorithm yields a matrix that is closer to the set of quantum channels than the original matrix, but it remains outside the set. Typically, for a matrix $Z'$ that is TP but no CP this issue is addressed by mixing $Z'$  with a matrix proportional to the identity at the end, as expressed by
\begin{align}
    Z= (1-p)Z' + p \,\mathbb{1}/d^2.
    \label{eq:correc-id}
\end{align}
Here, $p$ is the solution to $(1- p)\lambda_{\min}+ p/d^2 = 0$, where $\lambda_{\min}$ represents the smallest eigenvalue of $Z'$. Although the matrix $Z$ is a sub-optimal solution for the problem, it falls within the set of quantum channels.  

Our proposed approach involves conducting several iterations of Dykstra's algorithm, followed by correcting the resulting matrix using the CBA method, instead of mixing it with the identity matrix.
Indeed, each iteration of Dykstra's algorithm exploits projections onto convex sets, specifically the $TP$ and the density matrices sets, which we call $CP1$.  The solution to these projections are analytically known in the literature (see App.~\ref{app:analytical_projections}).
The output matrix from this step is a CP matrix, but in general not TP, precisely the required input for the CBA step.
The complete algorithm for finding the closest Choi state via Dykstra's projection algorithm and adding the CBA correction at the end is outlined in Algorithm 1.


\begin{algorithm}[H]
\begin{algorithmic}[1]
\Procedure{dykstraCBA}{$X$ initial state, $\mu$ max. iters, $\epsilon$ tolerance}
\State $p_0, q_0 = 0$
\State $k = 0$
\State $\epsilon' = 100$
\While {$ \epsilon' \geq \epsilon$ and $k  \leq \mu$}
    \State $Y \gets$ proj$_{TP}(X + p_k)$
    \State $p_k = X + p_{k-1} - Y$
    \State $X \gets$ proj$_{CP1}(Y + q_k)$
    \State $q_k = Y + q_{k-1} - X$
    \State $k \gets k + 1$
    \State $\epsilon' = ||p_{k} - p_{k-1}||_F^2 + ||q_{k} - q_{k-1}||_F^2$
\EndWhile
\State $X_A \gets {\rm tr}_S(X)$
\State $ U \gets \mathbb{1}_d \otimes (\sqrt{dX_A })^{-1}$
\State $J^* \gets U X U^\dagger$ \\
\Return $J^*$
\EndProcedure
\end{algorithmic}
\caption{Projection onto $CPTP$ }
\end{algorithm}

\subsubsection{Numerical simulations}\label{sec:simulations-choi}
To evaluate our method, we start by considering the input matrix as the Choi state of a random Haar unitary channel $\mathcal{U}$ corrupted by random noise.
Specifically, we define the modified Choi state as follows:
\begin{equation}
\tilde{J} := (1-p)J+pN, \qquad p\in[0,1],
\label{input-mat}
\end{equation}
where $J$ is the Choi matrix of the channel $\mathcal{U}$, and $N$ is an Hermitian matrix constructed as
\begin{equation}
\label{complex-noise}
N=\frac{M+M^\dag}{{\rm tr}(M+M^\dag)}, \qquad
M_{ij}=a_{ij} + i \, b_{ij},
\end{equation}
with $a_{ij}$ and $b_{ij}$ being random real numbers drawn from a standard normal distribution. The matrix in  Eq.~\eqref{input-mat} has trace one but is not guaranteed to be PSD, nor TP as long as $p>0$. For our simulations, we consider $p \in \{0.001, 0.01, 0.1\}$.

To test the performance of our algorithm, we compared our protocol with existing approaches commonly used for finding the closest Choi state. We considered the following five methods:
\begin{enumerate}[(i)]
\item Cholesky-based approximation (CBA): Projects the input matrix onto the set of density matrices $CP1$ of dimension $d^2\times d^2$, and subsequently applies the correction derived in Eq.~\eqref{eq:closest_choi_cholesky} to project it onto the trace-preserving set.
\item Two-stage solution (TSS)~\cite{xiao2022two}: Projects the input matrix onto the set of positive matrices $CP$ by setting to zero the negative eigenvalues, and projects the resulting matrix onto the $TP$ set using the same analytical projection as in (i).  
\item HIPSwitch~\cite{Surawy_Stepney_2022}: 
Alternates between Alternate Projections (AP) and Hyperplane Intersection Projection (HIP) to iteratively update the input matrix, ensuring it satisfies the CPTP conditions. AP mode alternates between CP1 and TP projections, while HIP mode accumulates half-spaces that contain CP1 that are kept if the projection onto the intersection between TP and the half-spaces or hyperplanes defined in AP is the same. Switching between the two is based on cosine similarity or step counts and the process continues until convergence. Finally, the resulting state is mixed with the identity matrix (see Eq.~\eqref{eq:correc-id}) to ensure positivity, as detailed in App.~\ref{appendix:HIP}.

\item Dykstra + CBA: It alternates between $CP1$ and $TP$ projections with intermediate steps. We run this adopting the stopping criteria proposed in Ref.~\cite{birgin2005robust}, and then the Cholesky-based projection is applied to the resulting matrix. Implementation of the pseudocode is shown in Algorithm 1.
\item Dykstra + Id: Same as the method above, but the final correction is performed by mixing the resulting matrix with the identity to ensure the result is in the set of quantum channels as in Eq.~\eqref{eq:correc-id}. 
\end{enumerate}

\begin{figure}[t]
     \begin{subfigure}{1\columnwidth}
         \includegraphics[width=\textwidth]{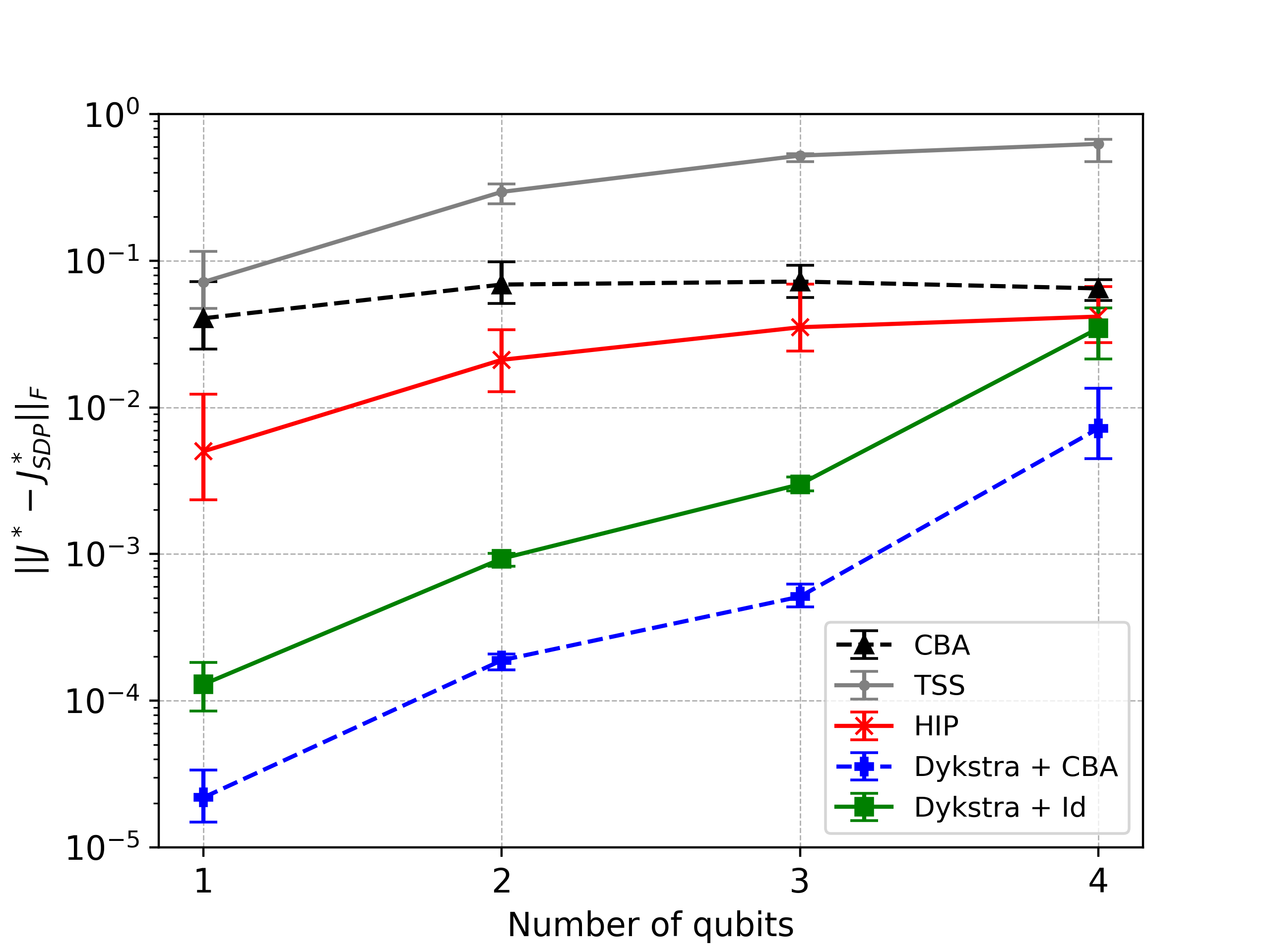}
         \caption{Frobenius distance to the exact Choi}
         \label{fig:precision_choi_p01}
     \end{subfigure}
     \begin{subfigure}{1\columnwidth}
         \includegraphics[width=\textwidth]{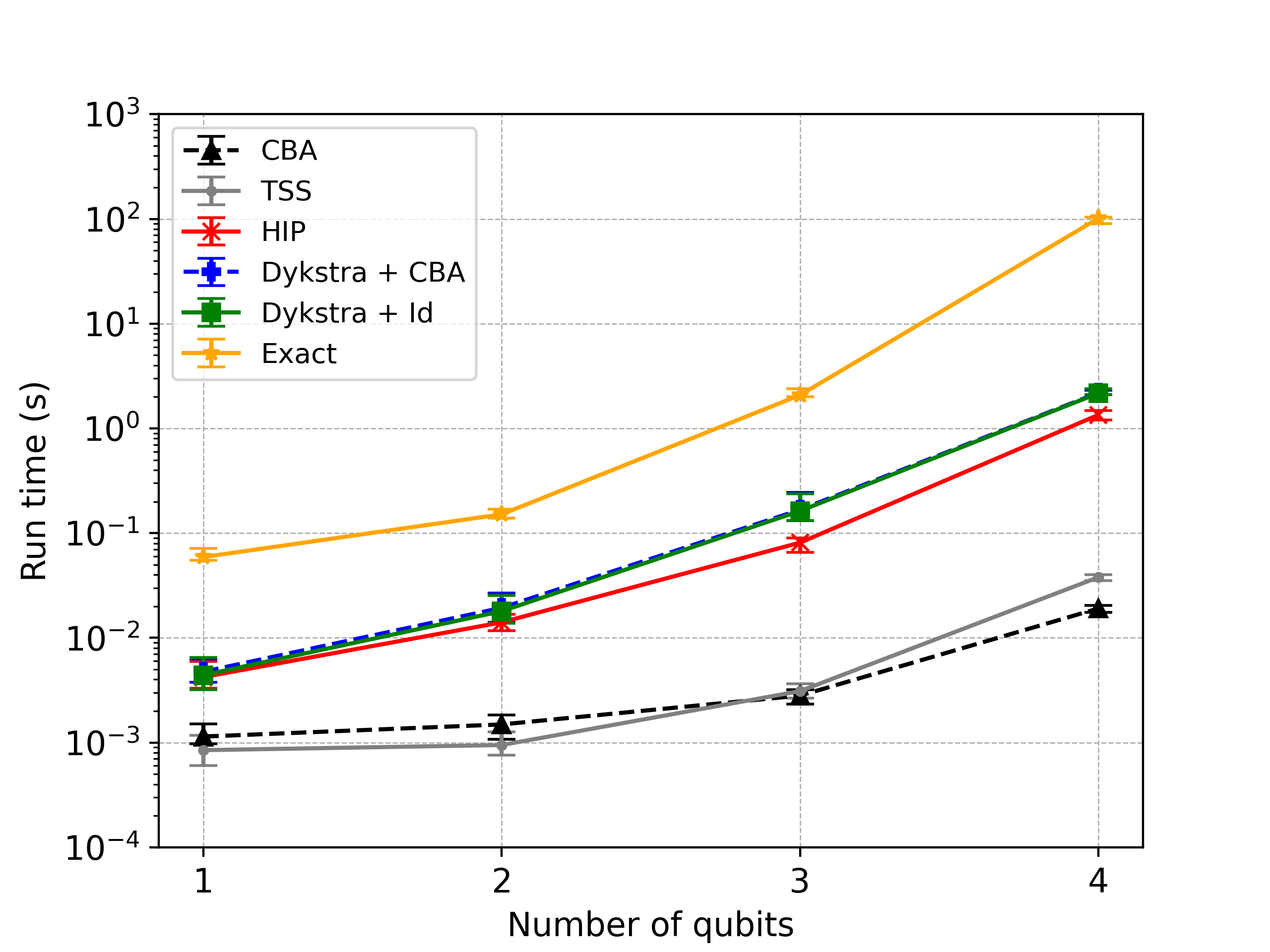}
         \caption{Running time to find the closest Choi}
         \label{fig:time_choi_p01}
     \end{subfigure}
     \caption{\justifying In Fig.~\ref{fig:precision_choi_p01}, we illustrate the Frobenius distance between the exact Choi state obtained by the semidefinite programming method (\texttt{SCS, eps =} $1.0\times 10^{-8}$) and the five considered methods (with precision $1.0\times 10^{-7}$), with our methods represented by dashed lines. Fig.~\ref{fig:time_choi_p01} depicts the computational time in seconds taken by the different methods compared to the SDP approach. The data reflect median values computed from 100 samples, with noise levels set to $p = 0.1$, with error bars indicating the interquartile range (IQR) around the median.}
      \label{fig:choi}
\end{figure}

In Fig.~\ref{fig:precision_choi_p01}, we depict the Frobenius distance between the exact result found by SDP and the state returned by the five different methods as a function of the number of qubits of the channel to which the Choi state is associated. The results were obtained by computing the median of the methods applied to 100 random input matrices generated as in Eq.~\eqref{input-mat} with $p=0.1$. In Fig.~\ref{fig:time_choi_p01}, we display the time taken by the different projection methods. For the simulations, we employed the stopping condition for Dykstra's algorithm outlined in App.~\ref{appendix:dykstra} with $\epsilon = 1.0\times 10^{-7}$. This precision parameter was also used to determine the stopping condition for the HIPSwitch algorithm based on the value of the minimum eigenvalue. The results for other noise levels ($p = 0.01, 0.001$) are provided and discussed in App.~\ref{app:more-simulations} as well as the results for 10 and 12-qubit channels.

From Fig.~\ref{fig:precision_choi_p01} we observe that the Dykstra's method with the identity mixing (in green) achieves nearly two orders of magnitude higher precision than the HIPSwitch algorithm for a 1-qubit channel. However, as the system size increases, the behavior of Dykstra's method becomes indistinguishable from that of the HIPSwitch algorithm (depicted in red). Regarding running time, both algorithms exhibit comparable performance, taking approximately the same time to execute.
Upon comparing these results with those obtained by Dykstra + CBA (in blue), we observe that while maintaining a constant runtime, we achieve significantly higher precision compared to the rest of the algorithms. Particularly noteworthy is the enhancement of $\approx 3$ orders of magnitude in precision compared to HIPSwitch from one-qubit to three-qubit channels, while for four-qubit channels this improvement is more moderate. This is generally a consequence of the stopping condition that has been set, limiting Dykstra's subroutine to perform the maximum number of iterations set to 100 which does not ensure reaching the threshold in precision (see Table~\ref{table:iterations-dykstra} in the App.~\ref{app:more-simulations} for the median number of iterations taken). 

It is also remarkable that our CBA algorithm achieves a precision slightly less than one order of magnitude lower than HIPSwitch when allowing for only one projection onto the $CP$ set (depicted by the black line) while running at a tenfold speed advantage over HIPSwitch. Furthermore, when compared to the recent algorithm TSS (in grey), it is observed that projecting onto the set of density matrices rather than solely onto the set of positive matrices yields results that are one order of magnitude more precise while running at the same speed. 


\subsection{Quantum Detector Tomography with Cholesky-based approximation}

We assume a set of $N$
complex matrices $\{\tilde{F}_n\}_{n=1}^N$, each of dimension $d \times d$, that is a non-physical estimation of a POVM $\{F_n\}_{n=1}^N$. The goal here is to find an approximate projection onto the set of POVMs. In other words, the objective is to find an approximate solution to problem~\eqref{eq:proj_povm}. This approximation will be represented by a set of matrices contained in the intersection of a set of $N$ positive semidefinite matrices and the set of all $N$ matrices that sum up to the identity. Therefore, we can adopt a similar approach to the previous method: initially projecting onto the PSD matrix space and subsequently projecting this updated set onto the set of POVMs.

The projection onto the space of $N$ PSD matrices reads 
\begin{equation}
\begin{array}{rrclcl}
\{X_{n}^{*}\}  = \mathrm{arg} \min_{\{ X_n\} } & \multicolumn{3}{l}{{ \sum_{n=1}^N \lVert \tilde{F}_n - X_n \rVert}_F^2}\\
\textrm{s.t.} 
& X_n \geq 0.\\
\end{array}
\end{equation}
The solution to this problem is a set of PSD matrices $\{ X_n^{*} \}_{n=1}^N$, where $X_n^{*}  = Pos(\tilde{F}_n)$ can be obtained by setting to zero the negative eigenvalues obtained from the spectral decomposition of the input matrix. The elements of this set can always be represented in terms of their Cholesky decomposition, namely $X_n^{*} = E_n E_n^\dagger $, where $E_n$ represents a lower triangular complex matrix.

Since we now have a set of matrices $\{ X_n^{*} \}_{n=1}^N$ that are positive, we can proceed with the second step of our method and solve the optimization problem
\begin{equation}\label{eq:proj_pos_POVM}
\begin{array}{rrclcl}
\{Y_{n}^{*}\}  = \mathrm{arg}\min_{\{ Y_n\} } & \multicolumn{3}{l}{{ \sum_{n=1}^N \lVert E_n - Y_n \rVert}_F^2}\\
\textrm{s.t.} 
& \sum_{n=1}^N Y_n Y_n^\dagger = \mathbb{1},
\end{array}
\end{equation}
where now we are minimizing the distance between the Cholesky matrices $\lVert E_n - Y_n \rVert_F^2$ instead of the distance $\lVert E_n E_n^\dagger - Y_n Y_n^\dagger \rVert_F^2$. To solve this problem analytically, we employ the method of Lagrange multipliers, where the only constraint is that the set of matrices spanning the solution sums up to the identity. Consequently, the Lagrangian takes the following form:
\begin{align}
    \mathcal{L} = \sum_{n=1}^N \lVert E_n - Y_n \rVert_F^2 - \mathrm{tr}\left(\Lambda \left(\sum_{n=1}^N Y_n Y_n^\dagger - \mathbb{1} \right)\right),
    \label{eq:povm-lagrangian}
\end{align}
where $\Lambda$ is a $d \times d$ Hermitian matrix. 

To solve problem \eqref{eq:proj_pos_POVM} we start by imposing the equation ${\partial \mathcal{L}}/{\partial \overline{Y}} = 0$. To calculate this, we use the results from Table 1 in Ref.~\cite{koor2023short}, and we obtain
\begin{align}
\frac{\partial \mathcal{L}}{\partial \overline{Y}} =  Y_n - E_n - \Lambda Y_n = 0,
\end{align}
from which we can express the solution in terms of the Lagrange multipliers
\begin{align}
Y_n = (\mathbb{1} - \Lambda)^{-1} E_n.
\end{align}
The Lagrange multipliers are determined by imposing the constraint $\sum_{n=1}^N X_n = \mathbb{1}$, leading to
\begin{align}
\mathbb{1} = \sum_{n=1}^{N} (\mathbb{1} - \Lambda)^{-1} E_n E_n^\dagger (\mathbb{1} - \Lambda^\dagger)^{-1}.
\end{align}
From this we obtain 
\begin{align}
(\mathbb{1} - \Lambda)^2 = \sum_{n=1}^{N} E_n E_n^\dagger,
\end{align}
and then, the elements of the POVM that are the solution to the problem using the CBA are given by
\begin{align}
Y_n^* Y_n^{*\dagger} = \left({\sqrt{\sum_{k=1}^N E_k E_k^\dag}}\right)^{-1} E_n E_n^\dagger \left({\sqrt{\sum_{k=1}^N E_k E_k^\dag}}\right)^{-1}.
\label{eq:correction-povm}
\end{align}

The previous solution can be improved by employing Dykstra's projection algorithm as a starting point, performing alternating projections between the sum-to-identity and the $CP$ sets (whose analytical projections are detailed in App.~\ref{app:analytical_projections}). This iterative process ensures convergence to the intersection of these sets in the asymptotic limit. For a finite number of iterations, we obtain a set of matrices that are not necessarily a POVM. Then, we apply the final correction described in Eq.~\eqref{eq:correction-povm} to these matrices, obtaining an upper bound for the distance of the closest POVM from the initial set of matrices: $\sum_{n=1}^N \lVert \tilde F_n - Y_n^* Y_n^*{}^\dag \rVert_F^2 \geq \sum_{n=1}^N \lVert \tilde F_n - Z_n^* \rVert_F^2$. The complete process is described in Algorithm 2.

Algorithm 2 is then intended to represent an alternative 
to the demanding 
SDP routines to approximate the solution of \eqref{eq:proj_povm}.
We notice that
a method faster than SDP approaches was already introduced in Ref.~\cite{wang2021two}, referred as the Two-Stage Estimation (TSE) algorithm.
While this method enables the identification of a set of matrices satisfying the POVM constraints, it does not guarantee to obtain the closest set to the input matrices. Interestingly, we can also improve the TSE method by first running a few iterations of Dykstra's projection algorithm and then applying TSE immediately after the final projection to the sum-to-identity set.
%
\begin{algorithm}[H]
\caption{Projection onto POVM set }
\begin{algorithmic}[1]
\Procedure{dykstraCBA}{$\{Y_n\}$ noisy POVM, $\mu$ max. iters, $\epsilon$ tolerance}
\State $p_n^{(0)}, q_n^{(0)} = 0$
\State $k = 0$
\State $ \epsilon' = 100$
\While {$\epsilon' \geq \epsilon$ and $k \leq \mu$}
    \State $\{X_n\} \gets$ proj$_{CP}(\{Y_n + q_{n}^{(k)}\})$
    \State $\{p_{n}^{(k)}\} = \{Y_n + p_{n}^{(k-1)} - X_n\}$
    \State $\{Y_n\} \gets$ proj$_{Id}(\{X_n + p_{n}^{(k)}\})$
    \State $\{q_{n}^{(k)}\} = \{X_n + q_{n}^{(k-1)} - Y_n\}$
    \State $k \gets k + 1$ 
     \State $\epsilon' = \sum_n ||p_{n}^{(k)} - p_{n}^{(k-1)}||_F^2 + ||q_{n}^{(k)} - q_{n}^{(k-1)}||_F^2$
\EndWhile
$ U \gets \left (\sqrt{\sum_{n = 1}^N X_n}\right )^{-1}$
\For {$n = 1,...,N$} 
\State $Y_n
\gets U X_n U^\dagger$ \\
\EndFor
\Return $\{Y_n\}_{n = 1}^N$
\EndProcedure
\end{algorithmic}
\end{algorithm}

\subsubsection{Numerical simulations}\label{sec:simulations-povm}
For the numerical simulations, we use the set of projectors given by the spectral decomposition of $m$-qubit Pauli strings $P$, where $P = \otimes_{i=1}^{m} \sigma_i$ and $\sigma_i = \sigma_X$, $\sigma_Y$, or $\sigma_Z$. Each set contains $2^m$ matrices, and we randomly generate $100$ of these sets. Then, for each set, we modify each projector $F_n$ in the set as follows:
\begin{align}
\tilde{F}_n = (1-p)F_n + p N,
\label{eq:input-noise-povm}
\end{align}
where $N$ is a Hermitian matrix constructed as in Eq.~\eqref{input-mat}, and $p$ is set to $0.001$ for the simulations (see App.~\ref{app:more-simulations} for other noise levels and higher-dimensional systems).

We benchmarked our methods, CBA and Dykstra + CBA, against the existing algorithm TSE and an improved version of it, Dykstra + TSE, using generated data. The approches considered can be summarized as follows:\begin{enumerate}[(i)]
\item Cholesky-based approximation (CBA): Projects the set of $N$ input matrix onto the set of PSD matrices by setting to zero the negative eigenvalues, and subsequently applies the correction derived in Eq.~\eqref{eq:correction-povm} to project it onto the sum-to-identity set.
\item Two-stage estimation (TSE)~\cite{wang2021two}: Projects the set of input matrices onto the sum-to-identity set, and then projects the resulting matrices onto the POVM set via a unitary transformation (see App.~\ref{app:tse}).  
\item Dykstra + CBA: It alternates between $PSD$ and sum-to-identity projections with intermediate steps. We run this adopting the stopping criteria proposed in Ref.~\cite{birgin2005robust}, and then the Cholesky-based projection is applied to the resulting set of matrices. Implementation of the pseudocode is shown in Algorithm 2.
\item Dykstra + TSE: Same as the method above, but the final correction is performed with the unitary transformation as described in (ii). 
\end{enumerate}

\begin{figure}[t]
     \begin{subfigure}{1\columnwidth}
         \includegraphics[width=\textwidth]{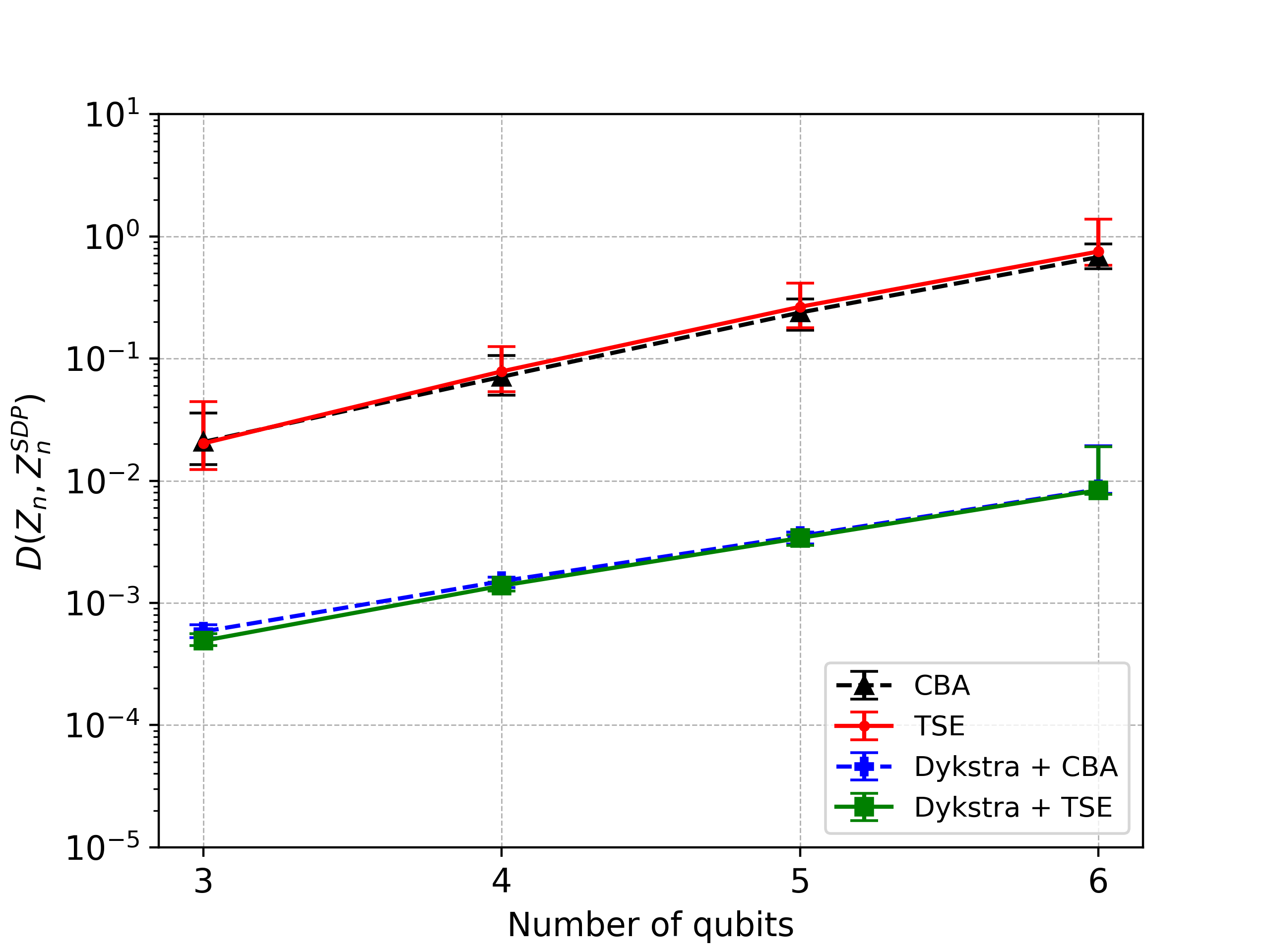}
         \caption{Frobenius distance to the exact POVM}
         \label{fig:precision_povm_p0001}
     \end{subfigure}
     \begin{subfigure}{1\columnwidth}
         \includegraphics[width=\textwidth]{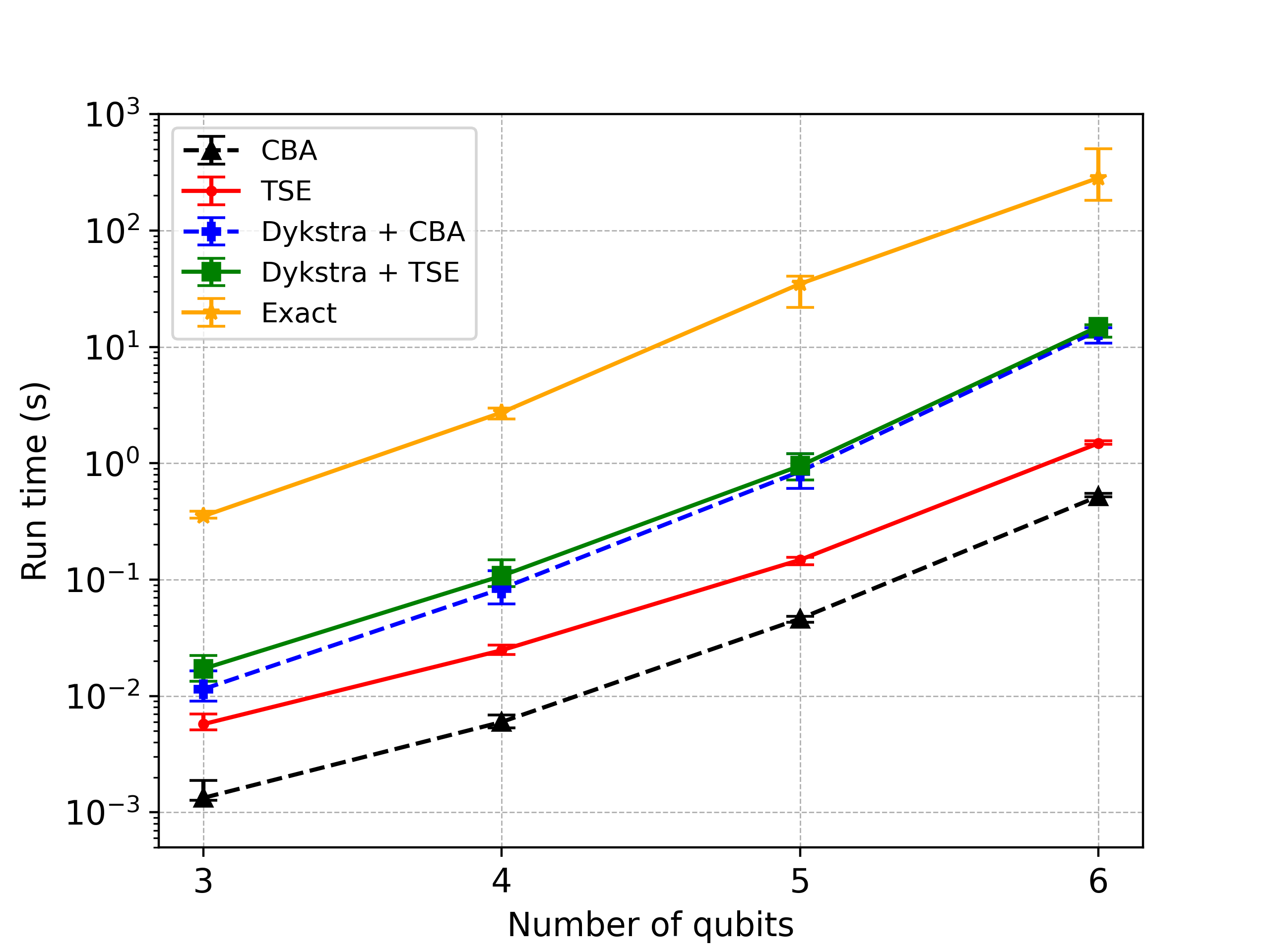}
         \caption{Running time}
         \label{fig:time_povm_p0001}
     \end{subfigure}
     \caption{\justifying In Fig.~\ref{fig:precision_povm_p0001}, we illustrate the Frobenius distance between the exact POVM obtained by the semidefinite programming method (\texttt{SCS, eps} = $1.0\times 10^{-8}$) and the five considered methods (with precision $1.0\times 10^{-7}$), with our methods represented by dashed lines. Fig.~\ref{fig:time_povm_p0001} depicts the computational time in seconds taken by the different methods compared to the SDP approach. The data reflect median values computed from 100 samples, with noise levels set to $p = 0.001$, with error bars indicating the interquartile range (IQR) around the median.}
      \label{fig:povm}
\end{figure}

In Fig. \ref{fig:precision_povm_p0001} we show the distance $D(Z_n, Z_n^{SDP}) = (\sum_{n=1}^N \lVert Z_n - Z_n^{SDP} \rVert_F^2)^{1/2}$ between the exact solution found by SDP and the result obtained when running CBA (black), TSE (red), Dykstra + CBA (blue), and Dykstra + TSE (green).  As the stopping condition for Dykstra's algorithm, we employ the criterion described in Ref.~\cite{birgin2005robust} with $\epsilon = 1.0\times 10^{-7}$. In Fig. \ref{fig:time_povm_p0001} we show the running time of the algorithms. 
From the results, we observe that CBA finds as accurate POVMs as the ones found by TSE, being approximately one order of magnitude faster than TSE and between two and three orders faster than the SDP. The precision of both CBA and TSE increases by up to two orders of magnitude when combined with Dykstra's projection algorithm, albeit at the expense of longer running times. Nonetheless, these running times remain significantly better than those of the SDP.

\section{Conclusions}
\label{sec:conc}
Recent methods have been proposed to tackle QPT and QDT more efficiently than traditional maximum likelihood estimation approaches, which can become computationally intractable for large numbers of variables. However, existing protocols often trade off computational performance for precision or vice versa.

We addressed the challenges of QPT by introducing a method for approximating projections onto sets of quantum channels. With this method, we can map a non-physical estimation of a quantum channel obtained using linear inversion techniques to a physical matrix. Our approach, based on the Cholesky decomposition, allowed us to derive an analytical expression for correction after an iterative projection scheme. When simulating our algorithm across 1 to 4-qubit channels, we observed that our method allows for addressing QPT with the same speed-up as current algorithms~\cite{Surawy_Stepney_2022}, while achieving precision that is three orders of magnitude greater. Additionally, we extended our algorithm to QDT, deriving an analytical expression for the closest POVM to a given set of positive matrices. Our analysis, conducted on 3 to 6-qubit systems, demonstrated that our algorithm outperforms current approaches for quantum detector tomography~\cite{wang2021two} in terms of both running time and precision.

%
\section*{Code availability}
Simulations were run using \texttt{Pyhton} 3.11.9 and \texttt{CVXPY} 1.4.1 on a Intel Core i5-7500 3.4 GHz 6 MB L3 cache. The computations could be optimized by utilizing a GPU or parallelization.
The scripts to run the experiments presented in this work can be found in GitHub in the repository \texttt{\href{https://github.com/juliabarbera/approximate_projections}{juliabarbera/approximate$\_$projections}}.

\begin{acknowledgments}
This work was supported by the ERC AdG CERQUTE, the AXA Chair in Quantum Information Science, the Government of Spain (Severo Ochoa CEX2019-000910-S, FUNQIP, European Union NextGenerationEU PRTR-C17.I1 and Quantum in Spain), the EU projects Veriqtas and PASQUANS2, Fundació Cellex, Fundació Mir-Puig, and Generalitat de Catalunya (CERCA program). JBR has received funding from the “Secretaria d’Universitats i Recerca del Departament de Recerca i Universitats de la Generalitat de Catalunya” under grant FI-1 00096, as well as the European Social Fund Plus.
DF acknowledges financial support from PNRR MUR Project No. PE0000023-NQSTI.
\end{acknowledgments}

\bibliography{bib}

\appendix
\section{Acronyms}\label{app:acronyms}
In this section, we provide a list of all the acronyms used throughout the manuscript in Tab.~\ref{table:acronyms}.
\begin{center}
\begin{table}[H]
\renewcommand{\arraystretch}{1.4}
\begin{tabular}{ |c|c|} 
 \hline
 Acronym & Description  \\ 
 \hline
  QST  & Quantum state tomography  \\ 
 \hline
 QPT &  Quantum process tomography \\ 
  \hline
  QDT & Quantum detector tomography\\
  \hline
  POVM & Positive operator-valued measure  \\ 
    \hline
    $CP$ &  Completely-positive set\\
    \hline
    $CP1$ & Set of density matrices\\
    \hline
    $TP$ &  Trace-preserving set \\
    \hline
    $CPTP$ & Completely-positive trace-preserving set\\
    \hline
    PSD & Positive semidefinite\\
    \hline
    CBA & Cholesky-based approximation\\
    \hline
    MLE & Maximum likelihood estimation \\
    \hline
    SDP & Semidefinite programming\\
    \hline
    TSS & Two-stage solution~\cite{xiao2022two} \\
    \hline
    HIP & Hyperplane intersection projection~\cite{Surawy_Stepney_2022}  \\
    \hline
    TSE & Two-stage estimation~\cite{wang2021two}  \\
    \hline
    AP & Alternate projections\\
    \hline
    LSR & Least-squares regression\\
    \hline
    IQR & Interquartile range \\
    \hline

\end{tabular}
\caption{Summary of the acronyms used in the manuscript.}
\label{table:acronyms}
\end{table}
\end{center}

\section{Dykstra's projection algorithm}\label{appendix:dykstra}

Dykstra's projection algorithm is a powerful method designed to determine the projection of a point onto the intersection of two or more closed convex sets.

Let $H$ denote a Hilbert space, with $C_A$ and $C_B$ being closed, convex subsets of $H$. Assume that, for any point $r$ in $H$, the projections onto each subset $P_{C_A} (r)$ and $P_{C_B} (r)$ can be calculated. Hence, the solution to the problem will be given by $x^\ast = \textrm{arg} \min_{x \in C_{A}\cap C_{B}} || x - r ||$. By Theorem~\ref{theorem:dykstra}, the projection onto the intersection $C_A \cap C_B$ can be obtained from the limit of the sequence ${x_k}$ defined by

\begin{align}\label{eq:dykstra-sequence}
y_k &= P_{C_A} (x_k + p_k)\nonumber\\
p_{k+1} &= x_k + p_k - y_k\nonumber\\
x_{k+1} &= P_{C_B} (y_k + q_k)\\
q_{k+1} &= y_k + q_k - x_{k+1}\nonumber,
\end{align}
with initial conditions $x_0 = r$, $q_0 = 0$, and $p_0 = 0$.
\begin{theorem}{(Boyle and Dykstra~\cite{boyle1986method}).} Let $C_{A}$ and $C_{B}$ be closed and convex sets of $\mathbb{R}_n$ such that $C=\cap_{i = A}^B C_i \neq \emptyset $. For any $i=A,B$ and any $x_0\in \mathbb{R}_n$, the sequence $\{x_k, y_k\}$ generated by~\eqref{eq:dykstra-sequence} converges to $x^\ast = P_{C}(x_0)$ (i.e.,$ |x_k - x^\ast| \rightarrow 0$ as $k \rightarrow \infty$).
\label{theorem:dykstra}
\end{theorem}

The convergence of the algorithm to the set of quantum channels is guaranteed by the robust stopping criteria proposed in Ref.~\cite{birgin2005robust}. For a given small tolerance $\epsilon$, the algorithm converges to the solution of the problem if the condition

\begin{equation}
c_I^k = \sum_{i = A}^B || y^{k-1}_i - y^{k}_i ||^2_F \leq \epsilon
\end{equation}
holds. Consequently, when utilizing Dykstra's subroutine, the stopping criterion is determined by

\begin{equation}
c_I^k = || p^{k-1} - p^{k} ||_F^2 + || q^{k-1} - q^{k} ||_F^2 \leq \epsilon.
\label{eq:stop-condition}
\end{equation}

\section{Algorithms for QPT}
\subsection{HIPSwitch}\label{appendix:HIP}

The \texttt{hyperplane$\_$intersection$\_$projection$\_$switch} function alternates between two projection methods— alternate projections (AP) and Hyperplane Intersection Projection (HIP) —to iteratively update the input, $\tilde{\Phi}$, ensuring it satisfies physical constraints such as being trace-preserving and completely positive~\cite{Surawy_Stepney_2022}. The switching between AP and HIP is governed by specific rules based on the progress of each projection mode. The main steps of the algorithm can be described as follows.

The algorithm begins with AP. In this mode, the input matrix $\tilde{\Phi}$ is projected onto the $CP1$ and $TP$ set iteratively. The projection continues in AP mode until a switching criterion is met. The switching from AP to HIP is determined by the  \texttt{alt$\_$to$\_$HIP$\_$switch} parameter, which can be set to one of two possible modes:
\begin{itemize}
    \item \textit{Cosine-based switching}: The algorithm computes the cosine similarity between successive projection vectors. If two consecutive steps become sufficiently collinear, indicated by the cosine of the vectors being at least \texttt{min$\_$cos}, the algorithm switches to HIP.
    \item \textit{Counter-based switching}: The algorithm tracks the number of AP steps performed. Once the number of steps reaches a predefined threshold (\texttt{alt$\_$steps}), the switch to HIP is triggered.
\end{itemize}

Once in HIP mode, the algorithm begins projecting onto the $CP1$ set, which ensures that the quantum state remains physically valid. In this mode, the projection is constrained by hyperplanes defined during the AP steps. The algorithm accumulates half-spaces that contain $CP1$ and iteratively adjusts the quantum state by incorporating these half-spaces into the projection if the projection onto the intersection of hyperplanes is equivalent to the projection onto the intersection of half-spaces. The duration of HIP mode is controlled by the \texttt{HIP$\_$to$\_$alt$\_$switch} parameter, via several switching strategies:
\begin{itemize}
\item \textit{First-based switching}: HIP continues until the first active hyperplane is discarded, enforcing the algorithm to return to AP.
\item \textit{Missing-based switching}: The algorithm tracks how many hyperplanes become irrelevant and stops HIP when a set number (\texttt{missing$\_$w}) are no longer needed.

    \item \textit{Part-based switching}: HIP stops when the contribution of the first hyperplane becomes too small, specifically if it drops below a fraction (\texttt{min$\_$part}) of the total step size.
    \item  \textit{Counter-based switching}: Similar to AP, HIP runs for a predefined number of steps ( \texttt{HIP$\_$steps}) before switching back to AP.
\end{itemize}

During each HIP step, the state is updated by projecting onto the intersection of the active hyperplanes (half-spaces). The algorithm also limits the number of hyperplanes it tracks through the  \texttt{max$\_$mem$\_$w} parameter, preventing memory overuse in high-dimensional systems. Each projection ensures that the quantum state stays within one of the sets by alternating between CP1 and TP constraints.

The algorithm stops when either the number of iterations reaches  \texttt{maxiter} or the state is sufficiently close to satisfying both the trace-preserving and completely positive conditions. The closeness is defined by the smallest eigenvalue of the output matrix $\Phi'$, which must be greater than the threshold  \texttt{least$\_$ev$\_$x$\_$dim2$\_$tol} divided by the dimension of the system.

To ensure that the final state ($\Phi$) satisfies the CPTP constraints, the output state of HIPSwitch ($\Phi'$) is mixed with the maximally mixed state such that 
\begin{equation}
    \Phi = (1-p)\Phi' + p\frac{\mathbb{1}}{d^2}.
    \label{eq:corr-HIP}
\end{equation}
The parameter $p$ can be found by solving $(1 - p)\lambda_{min} + p/d^2 = 0$, with $\lambda_{min}$ being the most negative eigenvalue of $\Phi'$.

\subsection{Two-Stage Solution for QPT}\label{app:tss}

The Two-Stage Solution (TSS)~\cite{xiao2022two} is a recent algorithm employed to solve the problem of QPT. It is composed of an estimation stage using Least-Squares Regression Estimation (LRE), and then a projection onto the set of quantum channels. The projection stage operates similarly to the CBA algorithm presented here, dividing the problem into two subproblems, each addressing one of the constraints.
To address the positivity constraint, TSS begins by computing the spectral decomposition of the Hermitian matrix associated with the input matrix. Afterwards, one can check if the singular values are negative and set these to zero. Consequently, the solution to the initial problem is obtained from its Hermitian part, yielding a matrix in the $CP$ set. This procedure is outlined in Ref.~\cite{higham1988computing}.

The second problem is addressed by employing an analytical projection scheme, which yields an outcome equivalent to that described in Eq.~\eqref{eq:closest_choi_cholesky}, albeit driven by a distinct approach.

\subsection{Two-Stage Estimation for QDT}\label{app:tse}
The algorithm for QDT outlined in Ref.~\cite{wang2021two} operates in two key stages. First, it estimates the POVM from measurement data using constrained least squares, ensuring the completeness condition is met. Since the resulting matrices may not always be physical, a refinement is required in stage two, where matrix transformations are applied.

The estimated POVM element, $\hat{E}_i$, is expressed as the difference between two positive semidefinite matrices: $\hat{E}_i = \hat{F}_i - \hat{G}_i$ with the optimal solution for $\hat{F}_i$ and $\hat{G}_i$ given in Ref.~\cite{wang2021two}. From the identity constraint, we have $\mathbb{1} = \sum_i \hat{E}_i = \sum_i \hat{F}_i - \sum_i \hat{G}_i$ that decomposes into $\mathbb{1}  + \sum_i \hat{G}_i= \hat{C}\hat{C}^\dagger $. Therefore,

\begin{equation}
     \sum_i \hat{C}^{-1}\hat{F}_i(\hat{C}^\dagger)^{-1} = \mathbb{1}.
\end{equation}
Taking into account that $\hat{C}\hat{C}^\dagger = \hat{C}\hat{U}\hat{U}^\dagger\hat{C}^\dagger$ for any unitary matrix $\hat{U}$, $\hat{U}^\dagger \hat{C}^{-1}\hat{F}_i(\hat{C}^\dagger)^{-1} \hat{U}$ can also represent a solution to the QDT problem. Stage two reduces to finding the optimal unitary matrix $\hat{U}$ that minimizes $||\hat{C}\hat{U} - \mathbb{1}||$, with the solution given by
\begin{equation}
    \hat{U} = \sqrt{\hat{C}^\dagger \hat{C}}\hat{C}^{-1}.
\end{equation}
Thus, the final POVM is $\hat{U}^\dagger \hat{C}^{-1}\hat{F}_i(\hat{C}^\dagger)^{-1}\hat{U}$.

In essence, the TSE algorithm first enforces the sum-to-identity constraint, followed by the projection onto the PSD cone to ensure a valid POVM. This is the opposite approach to our CBA algorithm, which applies different projections in a reversed order.

\section{Some analytical projections}\label{app:analytical_projections}

\subsection{$\mathbf{TP}$ Projection}
To project onto the $TP$ set of matrices one can employ analytical projection expressions provided in Ref.~\cite{Surawy_Stepney_2022}. The problem of projecting onto the $TP$ set can be expressed as the following optimization problem 
    \begin{equation}
    \begin{array}{rrclcl}
    \min_{X} & \multicolumn{3}{l}{{\lVert F - X \rVert}_F^2}\\
    \textrm{s.t.} 
    & {\rm tr}_S (X) = \mathbb{1}/d.
    \end{array}
    \end{equation}
Hence, the solution to this problem reads
    
\begin{equation}
    X = F + \frac{1}{d} \mathbb{1} \otimes \left(\frac{1}{d} \mathbb{1} - {\rm tr}_S (F) \right)
\end{equation}
as proven in Ref.~\cite{Surawy_Stepney_2022}.

\subsection{Identity Projection}
To project onto the space of matrices that sum one, equivalent to solving the problem
    \begin{equation}
    \begin{array}{rrclcl}
    \min_{X_n} & \multicolumn{3}{l}{{ \sum_{n=1}^N \lVert F_n - X_n \rVert}_F^2}\\
    \textrm{s.t.} 
    & \sum_{n=1}^N X_n = \mathbb{1},
    \end{array}
    \end{equation}
    one can proceed in a similar way and derive an analytical expression to compute these projections. Therefore, the solution to the problem reads~\cite{wang2021two}
    \begin{equation}
    X_n = F_n - \sum_{j=1}^N F_j/N + \mathbb{1}/N.
    \label{eq:proj-identity}
    \end{equation}

\subsection{$\mathbf{CP}$ Projection}
The projection onto the set of PSD matrices is well-known in the literature. One can essentially remove the negative eigenvalues after the eigendecomposition, while keeping the positive ones. 

Given a matrix $A$, its factorization can be written in terms of its eigendecomposition as $A = VDV^\dagger$. Then, the PSD matrix reads 
\begin{equation}
    B = VD'V^\dagger,
\end{equation}
where the diagonal matrix $D'$ is obtained by setting it equal to $D$ and replacing the negative eigenvalues with zero. 

\subsection{$\mathbf{CP1}$ Projection}
For projecting onto the set of $d\times d$-density matrices we use the algorithm proposed in Ref.~\cite{Surawy_Stepney_2022} which relies in a threshold projection process similar to the one introduced in Ref.~\cite{smolin2012efficient}, and runs in $O(d^3)$. Here, we outline the steps of the algorithm in~\cite{smolin2012efficient} which consists in a special case of the threshold projection process. We consider $\mu$ as the input matrix.

\begin{enumerate}
    \item Calculate the eigenvalues and eigenvectors of $\mu$.
Arrange the eigenvalues is order from largest to
smallest. Call these $\mu_i, \ket{\mu_i}, 1\leq i \leq d$.
\item Let $i = d$ and set an accumulator $a = 0$.
\item If $\mu_i + a/i$ is non-negative, go on to step 4. Otherwise, set $\lambda_i = 0$ and add $\mu_i$ to $a$. Reduce $i$ by 1
and repeat step 3.
\item Set $\lambda_j = \mu_j + a/i$ for all $j \leq i$.
\item Construct $\rho = \sum_i \lambda_i \ket{\lambda_i}\bra{\lambda_i}$.
\end{enumerate}
Notice that when the routine is applied to process tomography the complexity becomes $O(d^6)$ since the Choi operator is a $d^2 \times d^2$-matrix.

\section{Extended numerical simulations}\label{app:more-simulations}
\subsection{Different noise levels}
\begin{figure*}[t]
     \centering
     \begin{subfigure}{1\columnwidth}
         \centering
         \includegraphics[width=\textwidth]{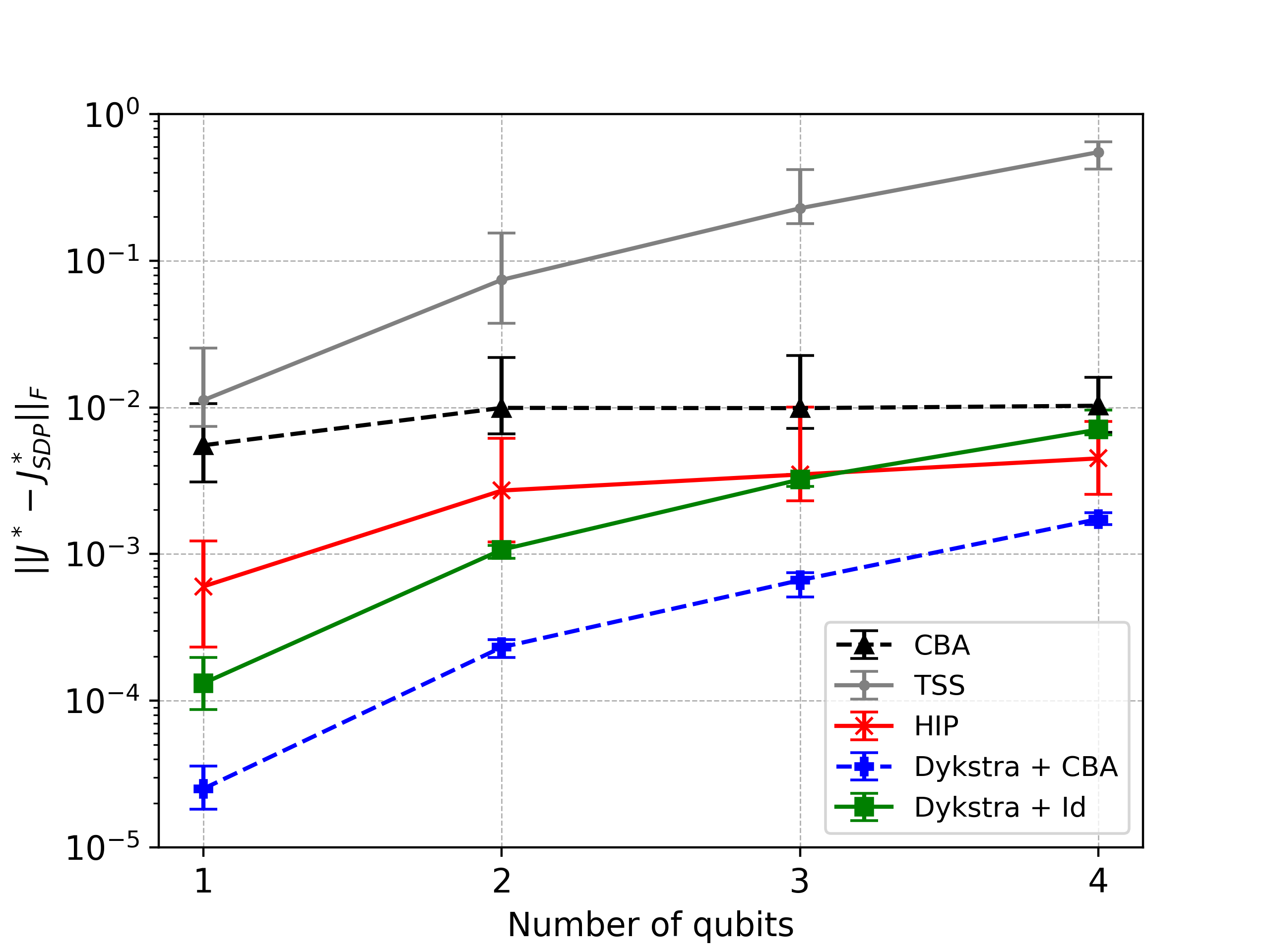}
         \caption{Frobenius distance to the exact Choi with $p = 0.01$}
         \label{fig:precision_choi_p1e2}
     \end{subfigure}
     \begin{subfigure}{1\columnwidth}
         \centering
         \includegraphics[width=\textwidth]{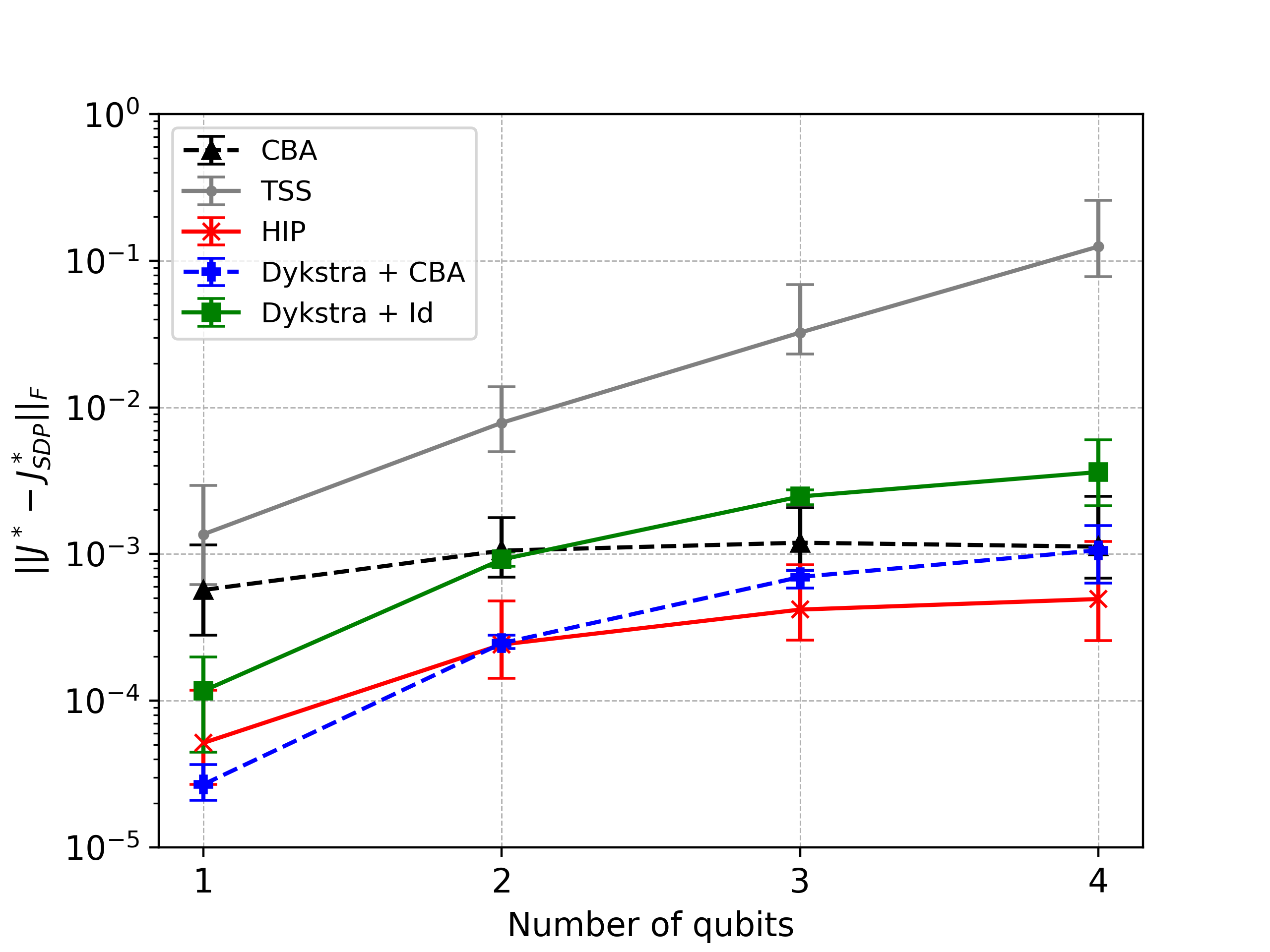}
         \caption{Frobenius distance to the exact Choi with $p = 0.001$}
         \label{fig:precision_choi_p1e3}
     \end{subfigure}
     \begin{subfigure}{1\columnwidth}
         \centering
         \includegraphics[width=\textwidth]{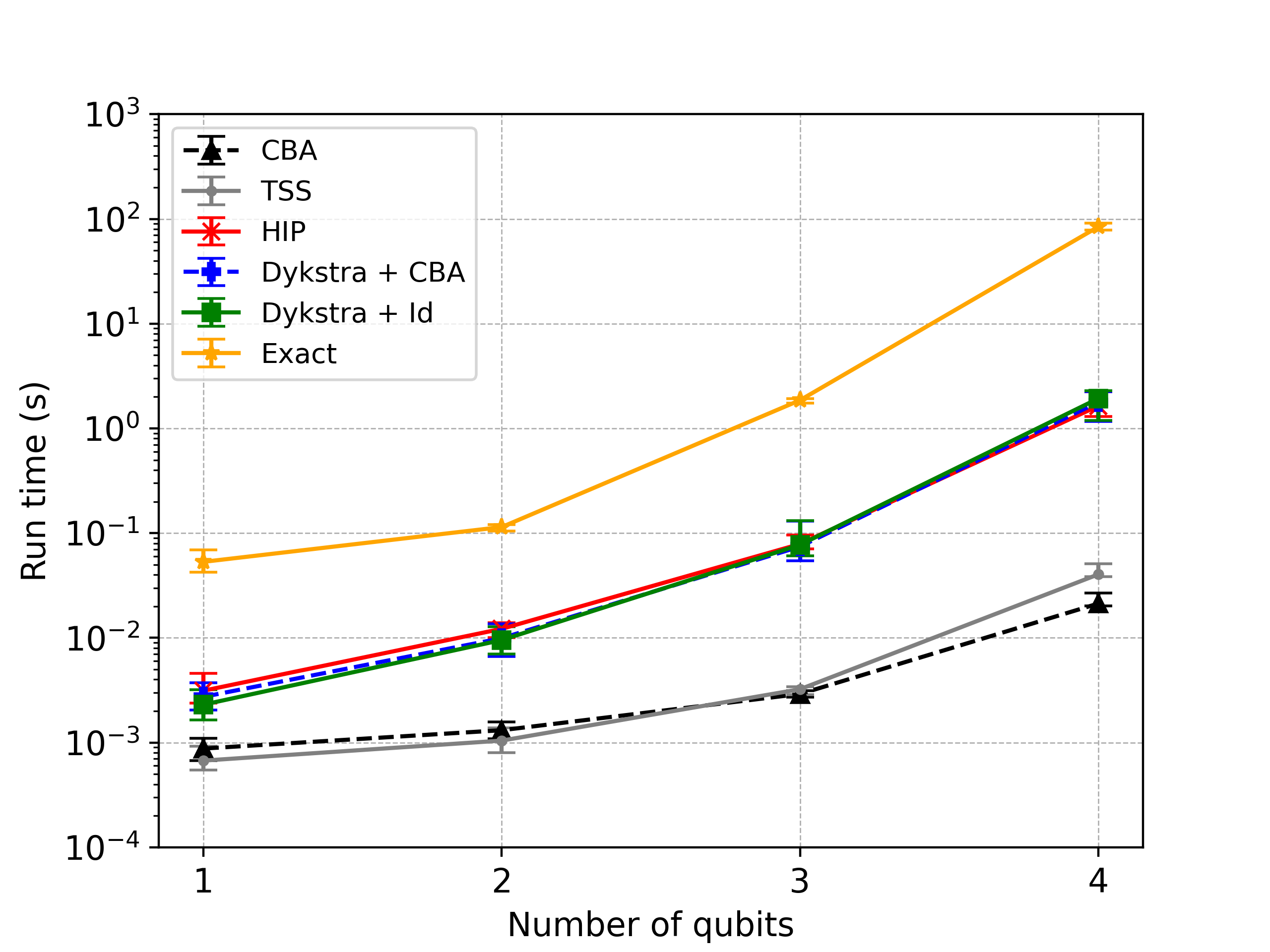}
         \caption{Running time to find the closest Choi with $p = 0.01$}
         \label{fig:time_choi_p1e2}
     \end{subfigure}
     \begin{subfigure}{1\columnwidth}
         \centering
         \includegraphics[width=\textwidth]{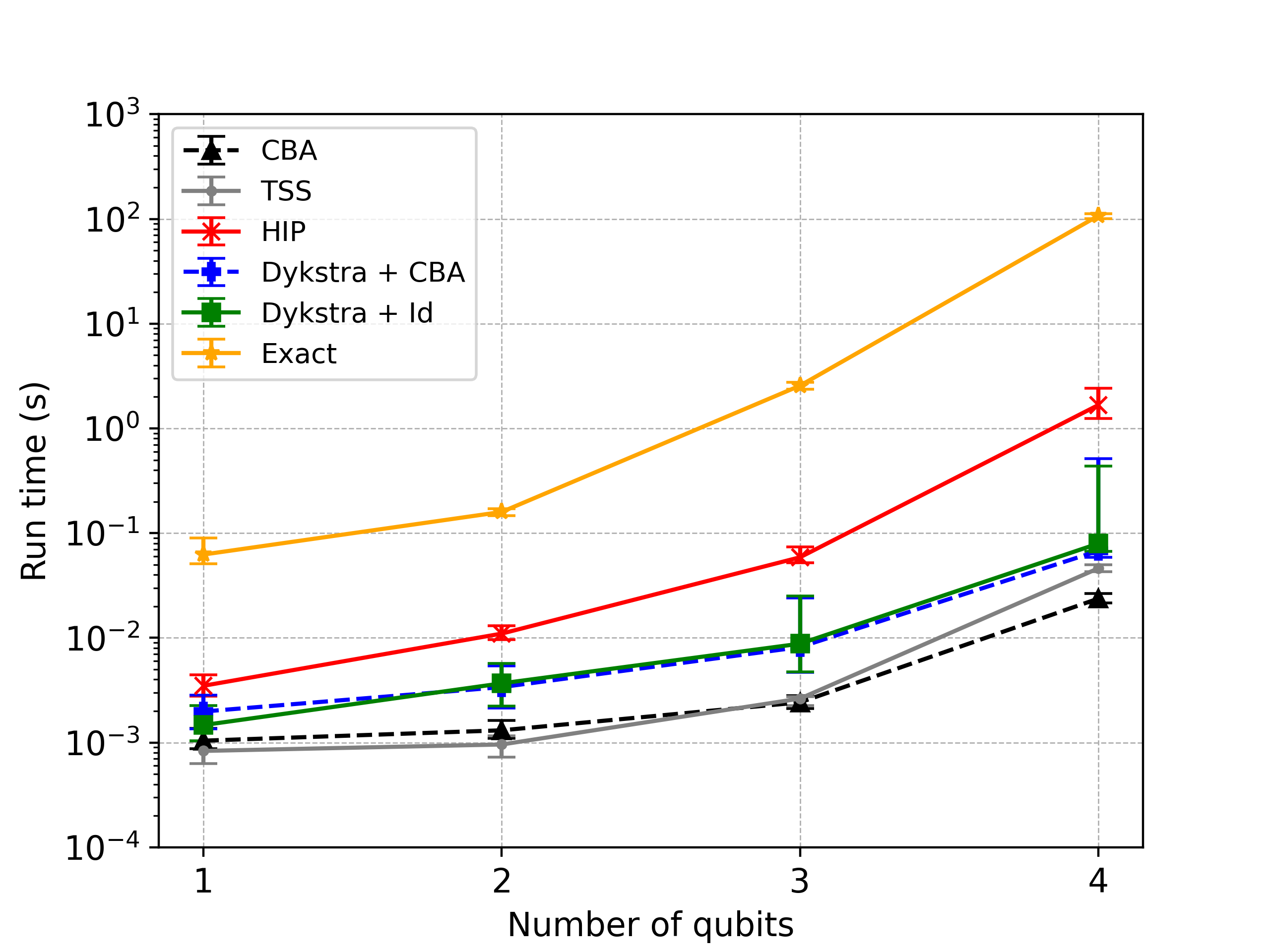}
         \caption{Running time to find the closest Choi with $p = 0.001$}
         \label{fig:time_choi_p1e3}
     \end{subfigure}
     \caption{\justifying In Fig.\ref{fig:precision_choi_p1e2} and Fig.\ref{fig:precision_choi_p1e3}, we illustrate the Frobenius distance between the exact Choi state obtained by the semidefinite programming method (\texttt{SCS, eps =} $1.0\times 10^{-8}$) and the five considered methods (with precision $1.0\times 10^{-7}$), with our methods represented by dashed lines. Fig.\ref{fig:time_choi_p1e2} and Fig.\ref{fig:time_choi_p1e3} depict the computational time in seconds taken by the different methods compared to the SDP approach. The data reflect median values computed from 100 samples, with noise levels set to $p = 0.01$ and $p = 0.001$, respectively, in each case, with error bars indicating the interquartile range (IQR) around the median.}
      \label{fig:choi-simulations-2}
\end{figure*}

 \begin{figure*}[t]
     \centering
     \begin{subfigure}{1\columnwidth}
         \centering
         \includegraphics[width=\textwidth]{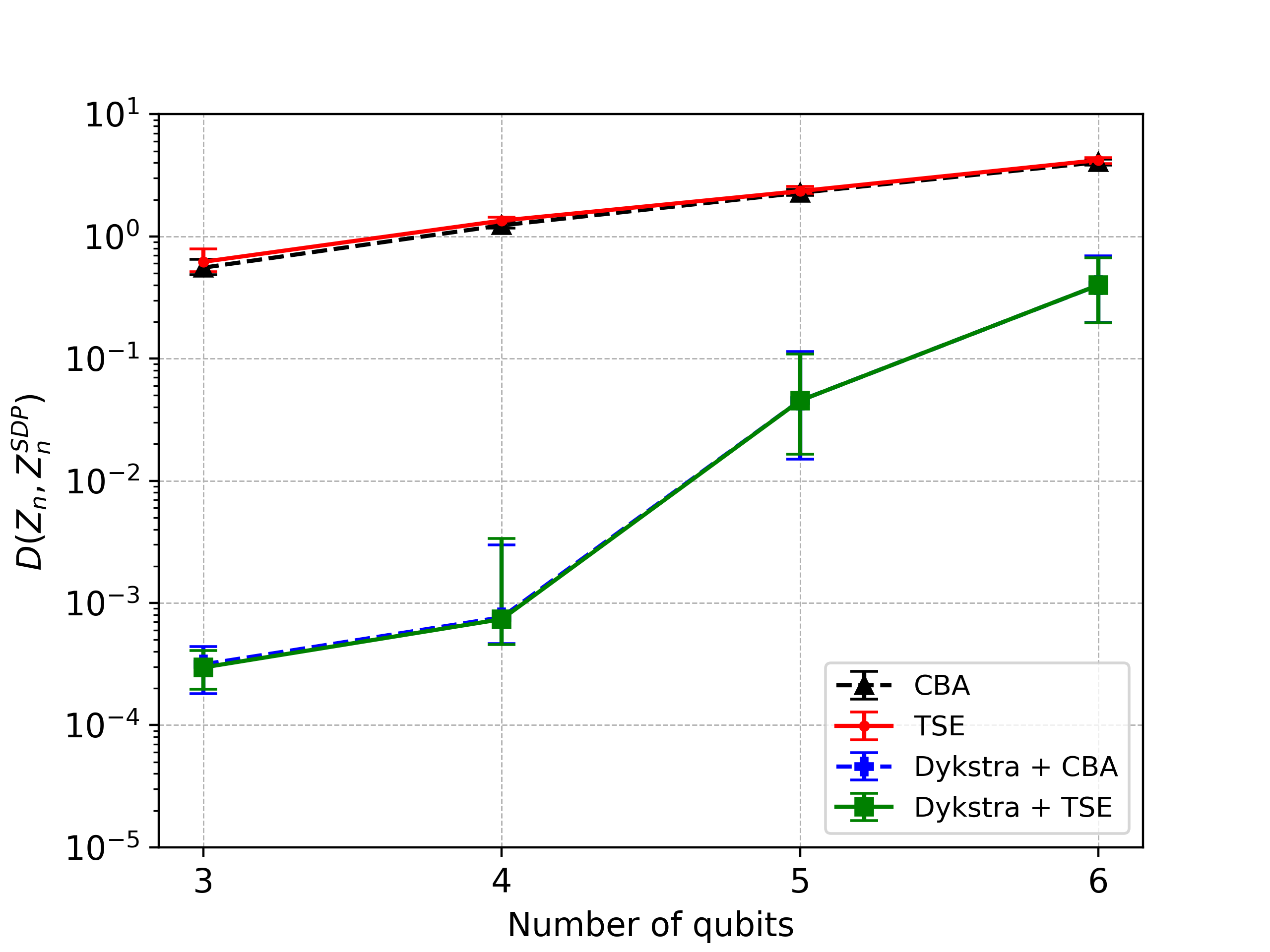}
         \caption{Frobenius distance to the exact POVM with $p = 0.1$}
         \label{fig:precision_povm_p01}
     \end{subfigure}
     \begin{subfigure}{1\columnwidth}
         \centering
         \includegraphics[width=\textwidth]{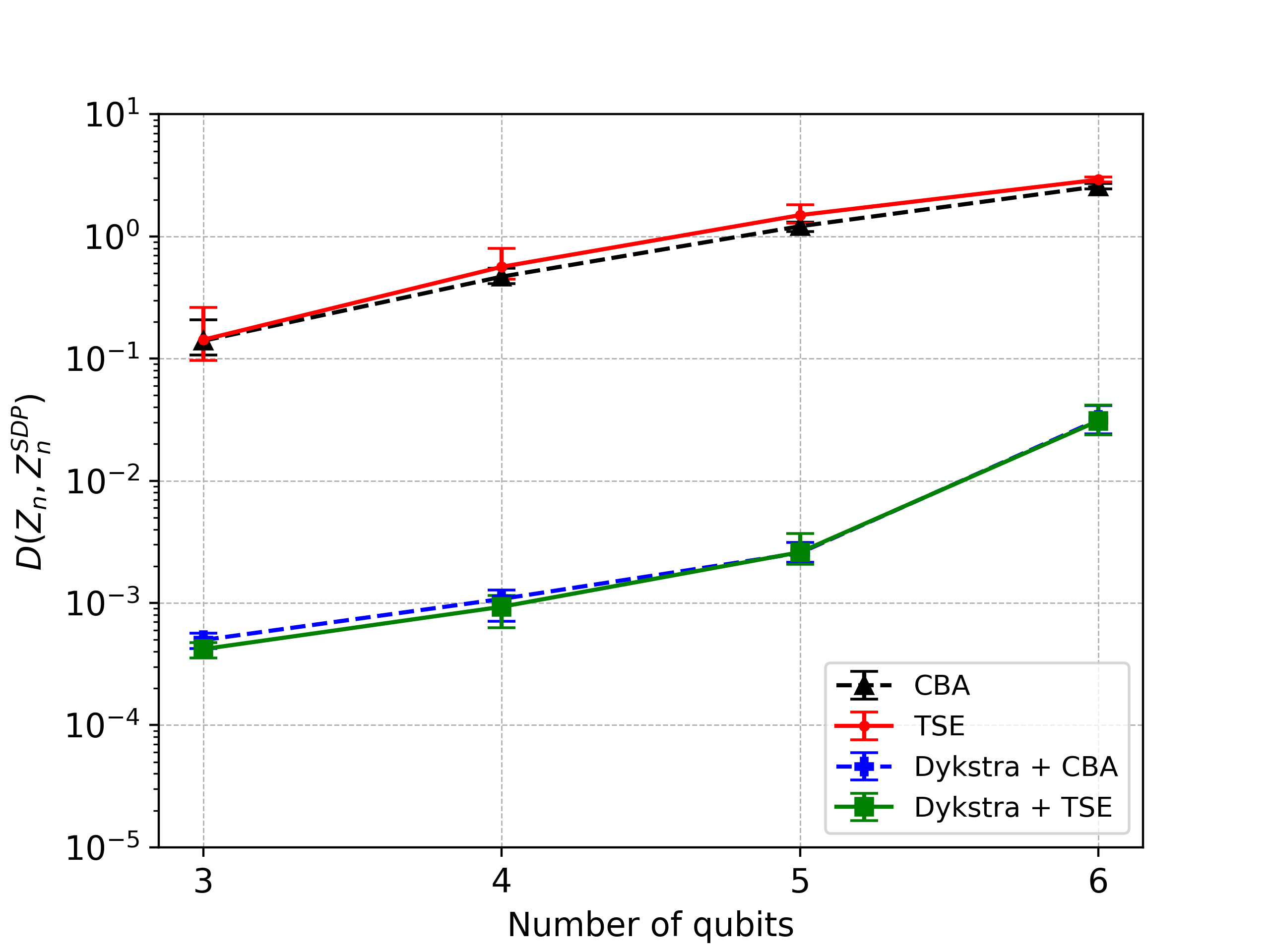}
         \caption{Frobenius distance to the exact POVM with $p = 0.01$}
         \label{fig:precision_povm_p001}
     \end{subfigure}
     \begin{subfigure}{1\columnwidth}
         \centering
         \includegraphics[width=\textwidth]{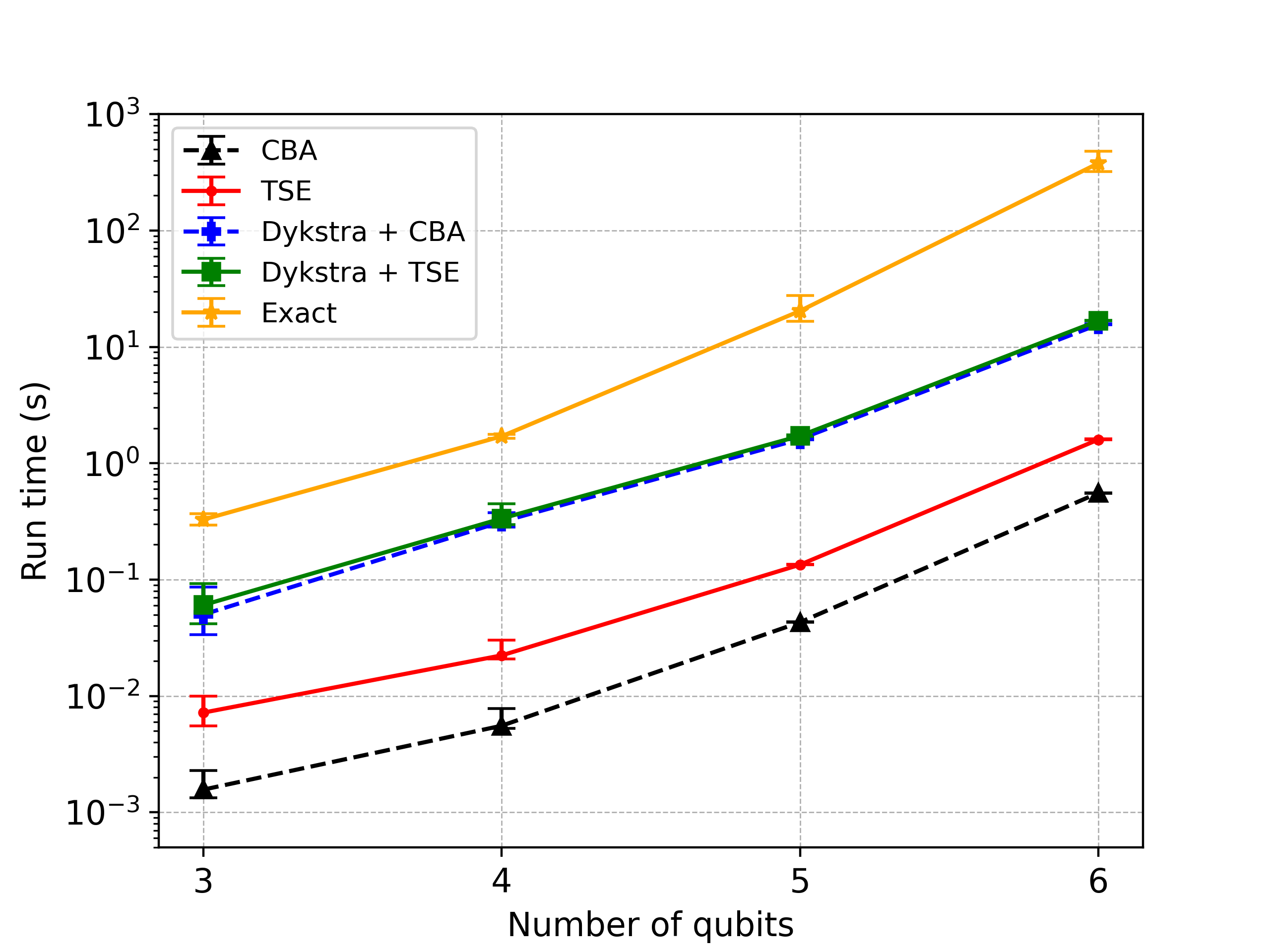}
         \caption{Running time taken to find the closest POVM with $p = 0.1$}
         \label{fig:time_povm_p01}
     \end{subfigure}
     \begin{subfigure}{1\columnwidth}
         \centering
         \includegraphics[width=\textwidth]{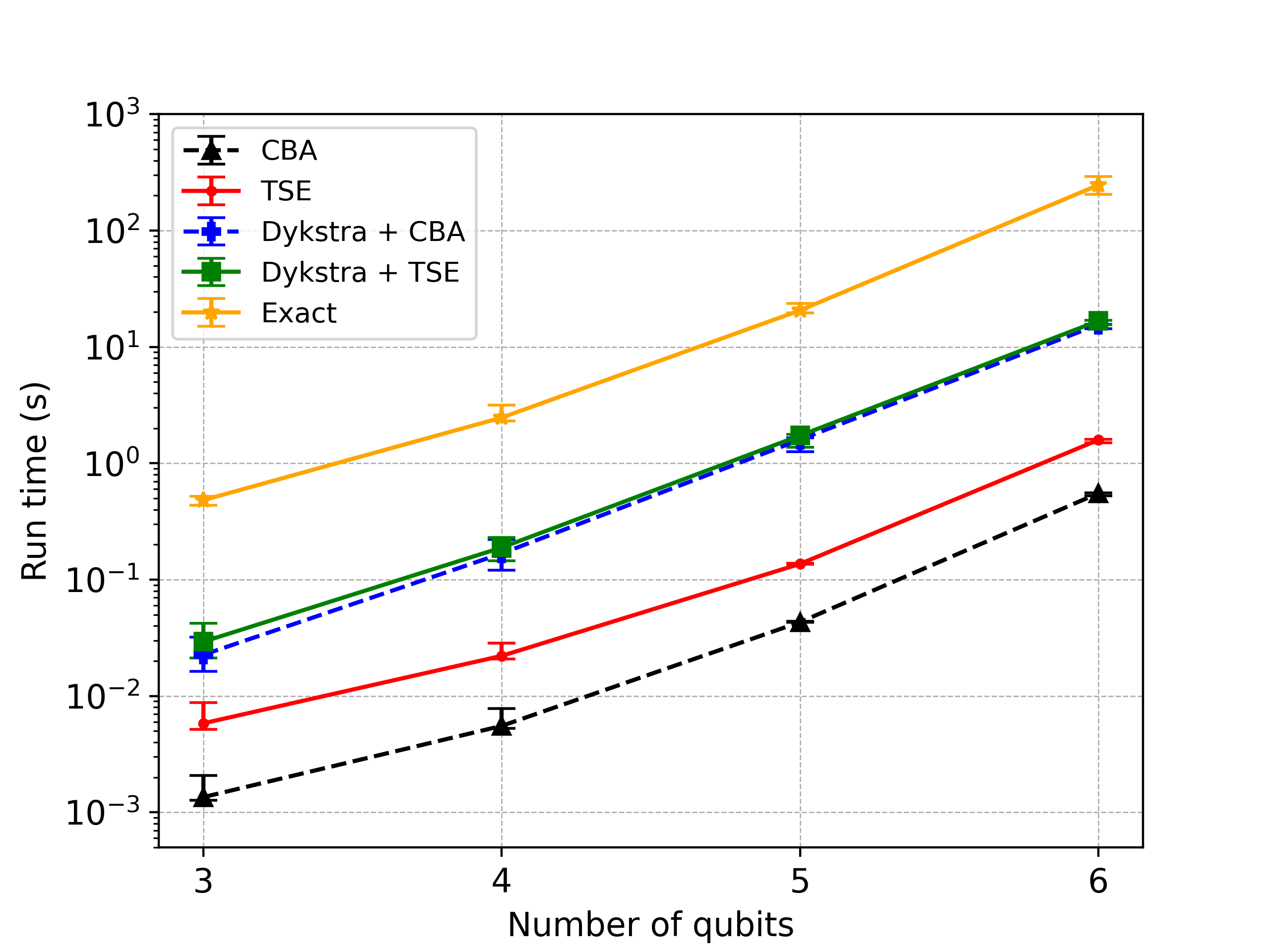}
         \caption{Running time taken to find the closest POVM with $p = 0.01$}
         \label{fig:time_povm_p001}
     \end{subfigure}
     \caption{\justifying In Fig.\ref{fig:precision_povm_p01} and Fig.\ref{fig:precision_povm_p001}, we illustrate the Frobenius distance between the exact POVM found by the semidefinite programming method (\texttt{SCS, eps =} $1.0\times 10^{-8}$) and the four considered methods (with precision $1.0\times 10^{-7}$), with our methods represented by dashed lines. Fig.\ref{fig:time_povm_p01} and Fig.\ref{fig:time_povm_p001} depict the computational time in seconds taken by the different methods compared to the SDP approach. The data reflect median values computed from 100 samples, with noise levels set to $p = 0.1$ and $p = 0.01$, respectively, in each case, with error bars indicating the interquartile range (IQR) around the median.}
      \label{fig:povm-simulations-2}
\end{figure*}

\begin{table*}
\begin{tabular}{l|llll|llll|}
\cline{2-9}
& \multicolumn{4}{c|}{Closest Choi}                         & \multicolumn{4}{c|}{Closest POVM}                         \\ \cline{2-9} 
& \multicolumn{1}{l|}{1 qubit} & \multicolumn{1}{l|}{2 qubits} & \multicolumn{1}{l|}{3 qubits} & 4 qubits & \multicolumn{1}{l|}{3 qubits} & \multicolumn{1}{l|}{4 qubits} & \multicolumn{1}{l|}{5 qubits} & 6 qubits \\ \hline
\multicolumn{1}{|l|}{$p$ = 0.001} & \multicolumn{1}{l|}{3.0}     & \multicolumn{1}{l|}{4.0}      & \multicolumn{1}{l|}{4.0}      & 2.0      & \multicolumn{1}{l|}{9.0}      & \multicolumn{1}{l|}{20.0}     & \multicolumn{1}{l|}{45.0}     & 93.5     \\ \hline
\multicolumn{1}{|l|}{$p$ = 0.01}  & \multicolumn{1}{l|}{5.0}     & \multicolumn{1}{l|}{13.0}     & \multicolumn{1}{l|}{30.0}     & 73.5     & \multicolumn{1}{l|}{17.0}     & \multicolumn{1}{l|}{40.0}     & \multicolumn{1}{l|}{93.0}     & 100.0    \\ \hline
\multicolumn{1}{|l|}{$p$ = 0.1}   & \multicolumn{1}{l|}{8.0}     & \multicolumn{1}{l|}{24.0}     & \multicolumn{1}{l|}{75.0}     & 100.0    & \multicolumn{1}{l|}{36.0}     & \multicolumn{1}{l|}{100.0}    & \multicolumn{1}{l|}{100.0}    & 100.0    \\ \hline
\end{tabular}
\caption{\justifying Median number of iterations required by Dykstra's algorithm to converge onto the set of quantum channels and onto the set of POVMs when considering the stopping condition presented in~\eqref{eq:stop-condition} with $\epsilon = 1.0\times 10^{-7}$. The analysis comprises 100 samples across different noise levels ($p = 0.1, 0.01, 0.001$) and input channel sizes ranging from 1 to 4 qubits for QPT and from 3 to 6-qubit systems for QDT.}\label{table:iterations-dykstra}
\end{table*}

Here, we present the results obtained from the experiments outlined in Sections~\ref{sec:simulations-choi} and~\ref{sec:simulations-povm} for noise levels $p \in [0.001, 0,01, 0.1]$. Table~\eqref{table:iterations-dykstra} displays the number of iterations necessary for Dykstra's algorithm to solve the QPT and QDT problems, satisfying the stopping condition in Eq.~\eqref{eq:stop-condition}, with a maximum of 100 iterations set.

The results in Fig.~\ref{fig:choi-simulations-2} reveal that the Dykstra + CBA algorithm exhibits improved running time performance as we reduce the noise of the input matrix, achieving a speed increase of one order of magnitude compared to recent algorithms. However, this enhancement in speed comes at the cost of slightly reduced precision, making it comparable to that of HIPSwitch.

For the POVM problem represented in Fig.~\ref{fig:povm-simulations-2} we observe that, as we increase the noise applied to the input POVM, the distance to the exact POVM increases when using TSE and CBA, while it remains constant for methods that combine with Dykstra's algorithm for low-qubit systems and worsens as we increase the system size. 
Moreover, as the system size increases, the computational time for all four methods gets worse such that they become only an order of magnitude faster compared to the SDP. This is expected because Dykstra's algorithm is halted after reaching the maximum number of iterations set to 100, limiting the possibility of achieving a better solution while consuming significant time.

\clearpage
\subsection{Higher-dimensional systems}

In this section, we also extend our numerical simulations to study 5-qubit channels for QPT and 7-qubit systems for QDT, assessing the scalability of our algorithms compared to existing methods. For systems of this size, obtaining the exact solution through SDP is infeasible with the available computational resources. Therefore, we compare the distance of the matrices obtained via projective methods from the Choi matrix and the POVM after noise corruption. 
Since all the methods we shall test provide upper bounds on the exact distance from the target set (e.g., for process tomography and method $X$ returning $J_X^*$, ${\Vert \tilde{J}- J_X^*\Vert}_F \geq {\Vert \tilde{J}-J^*  \Vert}_F $),
methods providing smaller values (for ${\Vert \tilde{J}- J_X^*\Vert}_F$) are expected to perform better in terms of precision.
To ensure manageable corruption levels as the dimensions increase, we set the noise level to $p = 0.0001$ for QPT and for QDT we reduce this noise level to $p = 0.00001$. We generated 100 samples for each method and specific instance, taking the median as a representative result. The maximum number of iterations for Dykstra's subroutine was set to 100, with a tolerance level of $\epsilon = 1.0\times 10^{-7}$.

In Tab.~\eqref{table:choi_metrics}, we present the distance to the Choi matrix after noise corruption and the time required to obtain each instance results for the algorithms considered. The results align with the trends observed for lower-dimensional channels: the CBA+Dykstra method achieves the highest precision without compromising computational efficiency. Notably, the CBA approach (the two-step algorithm proposed in this paper) yields as accurate results as its alternate version while requiring half of the runtime. It is also worth noting that the runtime of the HIP algorithm increases significantly with the dimension, making it less efficient compared to other methods that achieve comparable or superior precision. 

For QDT, we extend the results to 7-qubit systems, presenting both the distance to the noisy POVM and the runtime in Tab.~\eqref{table:povm_metrics}. Recall that, for lower-dimensional systems, the performance difference between the two-step methods CBA and TSE was minimal, though CBA had a slight runtime advantage. This behavior was similarly observed in alternate methods, which followed comparable trends.
For 7-qubit systems, however, our results indicate that the most accurate solutions are obtained with the Dykstra + TSE combination, followed by Dykstra + CBA. Notably, for systems of this size, Dykstra + TSE requires more than twice the runtime of Dykstra + CBA. Among the two-step methods, CBA yields more precise solutions than TSE, while operating in only half of TSE’s runtime.
\begin{table}[t]
\centering
\begin{subtable}[h]{0.8\linewidth} 
\centering
\begin{tabular}{|c|c|c|}
\hline
\multicolumn{3}{|c|}{Closest Choi (5-qubit channels)} \\ \hline
Method          & Distance      & Time (s)  \\ \hline
CBA             & 0.0045218     & 0.76      \\ \hline 
TSS             & 0.0580193     & 1.81      \\ \hline 
Dykstra CBA     & 0.0045217     & 1.66      \\ \hline 
HIP             & 0.0045241     & 76.38     \\ \hline 
Dykstra Id      & 0.0045605     & 3.09      \\ \hline
\end{tabular}
\caption{Closest Choi matrices for 5-qubit channels.}
\label{table:choi_metrics}
\end{subtable}

\vspace{0.1cm} 

\begin{subtable}[h]{0.8\linewidth} 
\centering
\begin{tabular}{|c|c|c|}
\hline
\multicolumn{3}{|c|}{Closest POVM (7-qubit system)} \\ \hline
Method          & Distance      & Time (s)  \\ \hline
CBA             & 0.0378        & 6.47      \\ \hline 
TSE             & 0.0402        & 14.19      \\  \hline
Dykstra CBA     & 0.0355        & 7.95      \\  \hline
Dykstra TSE     & 0.0296        & 18.06    \\ \hline
\end{tabular}
\caption{Closest POVM matrices for 7-qubit systems.}
\label{table:povm_metrics}
\end{subtable}

\caption{\justifying Median distances and computational times across 100 samples for (a) the Choi matrices of 5-qubit channels with  $p = 0.0001$ and (b) the POVM matrices for 7-qubit systems with  $p = 0.00001$. The tolerance for Dykstra was set to $\epsilon = 1.0\times 10^{-7}$.}
\label{table:combined_metrics}
\end{table}

We also evaluated our algorithms on 6-qubit channels in QPT (Choi states of 12 qubits). For QDT, we extended the evaluation to 8-qubit systems. For each case, we computed the median over 10 samples and used the same settings specified in the previous simulations. These simulations required additional computational resources, utilizing an Intel Xeon W-2295 3 GHz processor with 24.8 MB of L3 cache, without parallelization.

The results for QPT demonstrate that the CBA is already highly precise (time = 16.83 s., distance = 0.00834351), returning solution only $1.0\times 10^{-8}$ farther than its alternate version (time = 32.86 s., distance = 0.008343501), which is twice as slow. In comparison with the method HIP (time = 1656.58 s., distance = 0.00834507) proposed in Ref.~\cite{Surawy_Stepney_2022}, our CBA method is two orders of magnitude faster, while yielding Choi matrices that are only $1.0\times 10^{-6}$ farther from those obtained by HIP.
For QPT of 7-qubit channels, the computational cost becomes generally demanding, also in terms of memory requirements, so that we tested the several projective methods for one sample. Our results confirm that CBA (time = 774.48 s., distance = 0.0019778531) and Dykstra+CBA (time = 1475.09 s., distance =  0.0019778530, with two iterations) are faster and achieve comparable precision when compared with HIP (time = 76126.13 s., distance =  0.001979).

For QDT, our proposed Dykstra + TSE method  (time = 280.60 s., distance = 0.065) is only 1.20 times slower than the algorithm in Ref.~\cite{wang2021two} (time =  232.07 s., distanceTSE =  0.105), yet it is $4.0\times 10^{-2}$ more precise, yielding closer POVMs to the corrupted matrices. On the other hand, Dykstra + CBA (time = 88.66 s., distance =  0.077) shows achieves results that are $1.0\times 10^{-2}$ apart from the Dykstra + TSE results, but is three times faster. Lastly, the CBA algorithm proves to be slightly more precise than the current TSE method (time = 52.32 s., distance = 0.099) but is four times faster. 

These results highlight the computational efficiency of our methods relative to existing approaches, while providing solutions with comparable precision.

\end{document}